# Quantification of predictive uncertainty in hydrological modelling by harnessing the wisdom of the crowd: Methodology development and investigation using toy models

Georgia Papacharalampous[1,*], Demetris Koutsoyiannis[2], and Alberto Montanari[3]

[1] Department of Water Resources and Environmental Engineering, School of Civil Engineering, National Technical University of Athens, Heroon Polytechneiou 5, 157 80 Zographou, Greece; papacharalampous.georgia@gmail.com; https://orcid.org/0000-0001-5446-954X

[2] Department of Water Resources and Environmental Engineering, School of Civil Engineering, National Technical University of Athens, Heroon Polytechneiou 5, 157 80 Zographou, Greece; dk@itia.ntua.gr; https://orcid.org/0000-0002-6226-0241

[3] Department of Civil, Chemical, Environmental and Materials Engineering (DICAM), University of Bologna, via del Risorgimento 2, 40136 Bologna, Italy; alberto.montanari@unibo.it; https://orcid.org/0000-0001-7428-0410

* Correspondence: papacharalampous.georgia@gmail.com, tel: +30 69474 98589

**Abstract**: We introduce an ensemble learning post-processing methodology for probabilistic hydrological modelling. This methodology generates numerous point predictions by applying a single hydrological model, yet with different parameter values drawn from the respective simulated posterior distribution. We call these predictions "sister predictions". Each sister prediction extending in the period of interest is converted into a probabilistic prediction using information about the hydrological model's errors. This information is obtained from a preceding period for which observations are available, and is exploited using a flexible quantile regression model. All probabilistic predictions are finally combined via simple quantile averaging to produce the output probabilistic prediction. The idea is inspired by the ensemble learning methods originating from the machine learning literature. The proposed methodology offers larger robustness in performance than basic post-processing methodologies using a single hydrological point prediction. It is also empirically proven to "harness the wisdom of the crowd" in terms of average interval score, i.e., the obtained quantile predictions score no worse –usually better– than the average score of the combined individual predictions. This proof is provided within toy examples, which can be used for gaining insight on how the methodology works and under which conditions it can optimally convert point hydrological predictions to probabilistic ones. A large-scale hydrological application is made in a companion paper.

## 1. Introduction

Hydrological models are routinely applied for flood forecasting, water resources management and other environmental engineering applications (Montanari 2011). Their history, tracing back to 1850, can be found in Todini (2007), while their optimal design and exploitation (towards uncertainty reduction in hydrological modelling) remain since their early beginnings at the forefront of the hydrological research activity, currently consisting one of the only two modelling challenges included in the 23 major open problems in hydrology, as these problems were identified by Blöschl et al. (2019).

Based on their structure, hydrological models can be primarily classified as follows (see e.g., Solomatine and Wagener 2011; Pechlivanidis et al. 2011): (a) data-driven models, (b) conceptual models and (c) physically-based models. Models of categories (b) and (c) are also jointly called "process-based" (Montanari and Koutsoyiannis 2012). This specific term is largely associated with deterministic models by several authors (see e.g., Beven and Kirkby 1979; Makhlouf and Michel 1994; Perrin et al. 2003; Mouelhi et al. 2006a,b; Efstratiadis et al. 2008; Makropoulos et al. 2008; see also the applications by Madsen 2000; Nayak et al. 2013; Kaleris and Langousis 2017; Széles et al. 2018; Khatami et al. 2019, and the review by Efstratiadis and Koutsoyiannis 2010). On the contrary, models of category (a) are purely statistical. They are mostly borrowed from the statistical learning or machine learning literature (see e.g., Alpaydin 2010; Hastie et al. 2009; James et al. 2013; Witten et al. 2017) to be implemented in the hydrological literature with selected configurations and inputs (see e.g., Minns and Hall 1996; Dibike and Solomatine 2001; Solomatine and Dulal 2003; Nayak et al. 2013; Taormina and Chau 2015; Papacharalampous and Tyralis 2018; Tyralis and Papacharalampous 2018). In what follows, we use the terms "statistical learning" and "machine learning" interchangeably.

Data-driven and process-based models are, in fact, known to represent two different cultures or schools of thought in hydrological modelling, which need to be compromised in a way that will allow an optimal exploitation of predictability and uncertainty quantification (Todini 2007). In search of such a compromise, optimum (i.e., minimum error) point hydrological predictions (including forecasts) may result by post-processing the outcome of process-based models using statistical point prediction models (see e.g., Brath et al. 2002; Toth et al. 1999; Toth and Brath 2002; Abebe and Price 2003; Toth and Brath 2007). Hydrological post-processing methodologies aiming to convert point hydrological predictions, mostly predictions provided by process-based models, into probabilistic predictions are also available. These probabilistic methodologies utilize proper statistical models (i.e., probabilistic prediction or simulation models) complementary to the process-based ones. The statistical models used in post-processing are hereafter referred to under the term "error models", as they usually focus on the modelling of the hydrological model's error conditional on selected variables.

Here the interest is in probabilistic hydrological post-processing methodologies, in which the error model is estimated conditional upon the point prediction(s) of the hydrological model by using an independent segment (with respect to the one used for estimating the parameters of the hydrological model) extracted from the historical dataset. Various methodologies of this category are currently available (see e.g., Bock et al. 2018; Bourgin et al. 2015; Dogulu et al. 2015; Farmer and Vogel 2016; López López et al. 2014; Montanari and Brath 2004; Montanari and Grossi 2008; Montanari and Koutsoyiannis 2012; Solomatine and Shrestha 2009; Papacharalampous et al. 2019c; Tyralis et al. 2019b; Wani et al. 2017), amongst other probabilistic hydrological modelling and hydrological forecasting methodologies based on the idea of integrating process-based models and statistical approaches (see e.g., Beven and Binley 1992; Hernández-López and Francés 2017; Kavetski et al. 2002; 2006, Krzysztofowicz 1999, 2001, 2002; Krzysztofowicz and Kelly 2000, Krzysztofowicz and Herr 2001; Kuczera et al. 2006; Todini 2008; see also the review of Montanari 2011). Hereafter, we use the comprehensive term "two-stage" by Evin et al. (2014) to imply that the parameters of a probabilistic hydrological post-processing methodology are estimated within two subsequent stages.

Relying on the concept of ensemble simulations and opposed to "basic two-stage post-processing methodologies" utilizing a single point hydrological prediction (see e.g., Dogulu et al. 2015; Farmer and Vogel 2016; López López et al. 2014; Montanari and Brath 2004; Montanari and Grossi 2008; Papacharalampous et al. 2019c), the two-stage post-processing methodology by Montanari and Koutsoyiannis (2012) (hereafter referred to as "MK blueprint methodology") generates a large number of point hydrological predictions by using a single hydrological model (in its basic form; with different parameter values and ensemble inputs). These point predictions are hereafter referred to as "sister predictions" using the terminology of Nowotarski et al. (2016), Wang et al. (2016), and Liu et al. (2017). Different variants of the MK blueprint methodology can be found in Sikorska et al. (2015) and Quilty et al. (2019). The flexibility of the MK blueprint methodology is proved by the latter study, which focuses on probabilistic water demand forecasting using exogenous variables. Its main objective is converting point water demand forecasts produced by machine learning algorithms into probabilistic forecasts.

Here we introduce three novel variants of the MK blueprint methodology. These variants (hereafter collectively referred to as "methodology of the study") are inspired by the ensemble learning methods originating from the machine learning literature, while they are based on the concept of combining probabilistic predictions via simple quantile averaging from the forecasting field. Simple averaging (or equally weighted averaging or averaging) is a special form of linear combination (or linear pooling or weighted averaging) of predictions, in which all weights are equal (see e.g., Granger 1989; Winkler 2015; Wallis 2011; Lichtendahl et al. 2013). According to Granger (1989), (point) prediction combination can be traced back in the study of Barnard (1963), in which two point forecasts were averaged to form an outperforming forecast. Although having its roots in 1963 and more sophisticated combination approaches have been developed since then, this combination in simple fashion is even today suggested by Winkler (2015; see also Lichtendahl et al. 2013), because of its:

o   Interpretability.

- Simplicity in modelling.
- Better performance than weighted linear (or other) combinations in many cases.

In fact, as it is quoted from O'Hagan et al. (2006, p 190) in Lichtendahl et al. (2013), "simple, equally weighted opinion pool is hard to beat in practice". Moreover, it is the most common way of combining point or probability distribution function (PDF) forecasts (Lichtendahl et al. 2013; Wallis 2011). Especially when we are interested in combining a large number of predictions, as it is the case herein, simple averaging is rather the only reasonable option, also reminding us of several ensemble learning methods (see Hastie et al. 2009), e.g., the bagging by Breiman (1996) and random forests by Breiman (2001a), originating from the machine learning literature. These two examples of ensemble learning methods produce a large number of individual predictions and compute their average to finally produce the output prediction. This averaging leads to more accurate predictions, as it reduces their variance (Hastie et al. 2009, pp 282–288). Similarly, the average of quantile predictions may offer stability in performance, among other advantages. Quantile averaging has the following distinguishing features (Lichtendahl et al. 2013, Section 5; see also the interpretations provided by Winkler 2015):

- Under specific conditions (see e.g., the stylized versions examined in Lichtendahl et al. 2013) a predictor based on quantile averaging is robust. The same applies to a predictor based on PDF averaging.
- Under specific conditions (see e.g., the stylized versions examined in Lichtendahl et al. 2013) the average of quantile predictions scores no worse –usually better– than the average of scores of the combined individual predictions. This property (also applying to PDF averaging) is referred to as "ability to harness the wisdom of the crowd". Still, it has to be empirically proven for the problem and scores of interest.
- Quantile averaging can be convenient in practice, in contrast to PDF averaging.
- Quantile averaging is as useful as (or even more useful than) PDF averaging.

The methodology of the study has been developed in light of the above by also conducting a set of toy experiments (see e.g., Hartmann 1995; Frigg and Hartmann 2006; Klein and Romero 2007; Goldfarb and Ratner 2008; Luczak 2017; Reutlinger et al. 2017). Examples of toy experiments from the probabilistic hydrological modelling literature are available in Krzysztofowicz (1999), Beven and Freer (2001), Stedinger et al. (2008), Farmer and Vogel (2016), and Volpi et al. (2017). Toy models have also been exploited for other modelling situations in geoscience (see e.g., Koutsoyiannis 2006, 2010; see also the references in Koutsoyiannis 2006), while falling into the broader category of simulation or synthetic experiments, which are increasingly conducted within various hydrological contexts, including some more relevant to the present study (see e.g., Kavetski et al. 2002; Vrugt et al. 2005; Montanari 2005; Montanari and Koutsoyiannis 2012; Montanari and Di Baldassarre 2013; Papacharalampous et al. 2018, 2019a; Renard et al. 2010; Sadegh and Vrugt 2014; Sadegh et al. 2015; Sikorska et al. 2015; Schoups and Vrugt 2010; Tyralis and Papacharalampous 2017; Tyralis et al. 2013; Vrugt et al. 2003; Vrugt and Robinson 2007; Vrugt et al. 2008; Vrugt et al. 2013; Vrugt 2016); see also Montanari (2007) for a discussion on the significance of this type of experiments. In fact, simplified modelling situations can be useful as starting points for achieving effective real-world modelling, especially in cases where analytical solutions exist (see e.g., Volpi 2012).

The aims of the study are to:

1) Introduce a two-stage probabilistic hydrological modelling methodology that exploits in an optimal way (from a predictive modelling perspective) key concepts of the MK blueprint methodology.

2) Inspect the performance of the methodology of the study under known conditions and demonstrate how it works. In particular, we aim at testing whether and under which conditions this methodology can optimally convert point hydrological predictions to probabilistic ones.

3) Illustrate in simple fashion why and when it is meaningful for someone to select the methodology of the study over basic two-stage post-processing methodologies.

4) Increase the understanding on two-stage hydrological post-processing.

As implied by aims 2–4 and made e.g., by Krzysztofowicz (1999) and Stedinger et al. (2008), we herein present toy examples only. A companion paper by Papacharalampous et al. (2019b) is devoted to the validation of the methodology of the study using real-world data. In particular, in this latter paper we address a different set of research questions by conducting a large-scale experiment at monthly time scale. This experiment comprises 270 rainfall-runoff problems, which are found to be well-solved by the methodology of the study, while the larger robustness in performance of this methodology compared to basic two-stage post-processing methodologies is illustrated for all the examined problems. In the same experiment, we also clearly demonstrate the ability of the methodology of the study to harness the wisdom of the crowd.

## 2. An ensemble methodology for probabilistic hydrological modelling

In this section, we introduce a new methodology for probabilistic hydrological modelling, inspired by the MK blueprint methodology on the one hand and ensemble learning methodologies (see e.g., the review by Sagi and Rokach 2018) on the other hand. In what follows, random variables are underscored, following the Dutch convention.

### 2.1 Methodological background on two-stage post-processing

Two-stage post-processing methodologies are implemented by dividing the historical dataset into two independent segments. To outline the main steps and concepts adopted within a basic two-stage hydrological post-processing framework, we first define the time period $T = \{1, ..., (n_1+n_2+n_3)\}$, and its three distinct sub-periods $T_1 = \{1, ..., n_1\}$, $T_2 = \{(n_1+1), ..., (n_1+n_2)\}$ and $T_3 = \{(n_1+n_2+1), ..., (n_1+n_2+n_3)\}$. Let us now assume a historical rainfall-runoff dataset extending in the period $\{T_1, T_2\}$. Let us also assume that a probabilistic hydrological prediction is needed for the period $T_3$. Then the first segment of the historical dataset, extending in the period $T_1$, is used for calibrating the hydrological model, while information from the period $T_2$ is used to (a) apply the calibrated hydrological model, and (b) train (or fit) a suitable error model using the point prediction resulted from step (a) alongside with its target values. Under the stationarity and ergodicity assumptions (see e.g., Koutsoyiannis and Montanari 2015 for the implications of these assumptions in hydrological contexts), the trained error model can then be applied in the period $T_3$ for converting a point hydrological prediction obtained using the same hydrological model with the same parameters into a probabilistic hydrological prediction. The error model could fall into the category of conditional distribution models (see e.g., Montanari and Brath 2004; Montanari and Grossi 2008) or the category of statistical learning regression models that can directly provide predictive quantiles instead of predictive PDFs (see e.g., Dogulu et al. 2015; López López et al. 2014; Papacharalampous et al. 2019c; Tyralis et al. 2019b), amongst other model categories.

A summary of the MK blueprint methodology is also essential for what follows. As emphasized in Section 1, this two-stage post-processing methodology exploits information from a large number $m$ of sister predictions. Each sister prediction is obtained by utilizing the same hydrological model, yet with different parameter values and input data. The hydrological model's parameters are obtained by using data from the period $T_1$, while modelling and explicitly considering input data uncertainty imply the availability of input data error information. Information about the hydrological model's error, obtained from the period $T_2$, is then used to convert the sister predictions for the period $T_3$ to ensemble simulations of the process of interest. The $m$ ensemble simulations are retained as potential realizations of the process of interest, thus collectively composing a probabilistic prediction. For instance, if we are interested in delivering the 90% prediction interval and $m$ = 1 000, then we simply have to pick at each time $t \in T_3$ the 50$^{th}$ and 950$^{th}$ highest values (resulted via ranking) from the spaghetti plot of the 1 000 retained simulations. In absence of relevant information, the MK blueprint methodology can also be applied without explicitly considering input data uncertainty, i.e., by not running ensemble simulations for the hydrological model's input, without any loss of its generality (see e.g., the implementations in Quilty et al. 2019).

## 2.2 Methodology of the study (with three variants)

In this section, we present the methodology of the study. The presentation is made in a more formal and systematic manner with respect to Section 2.1, in which we summarize the methodological background of the study. Therefore, we here also set the largest part of the notations used throughout the paper. The formal presentation is accompanied by Figure 1, which summarizes in a compact way the methodological contribution of the study.

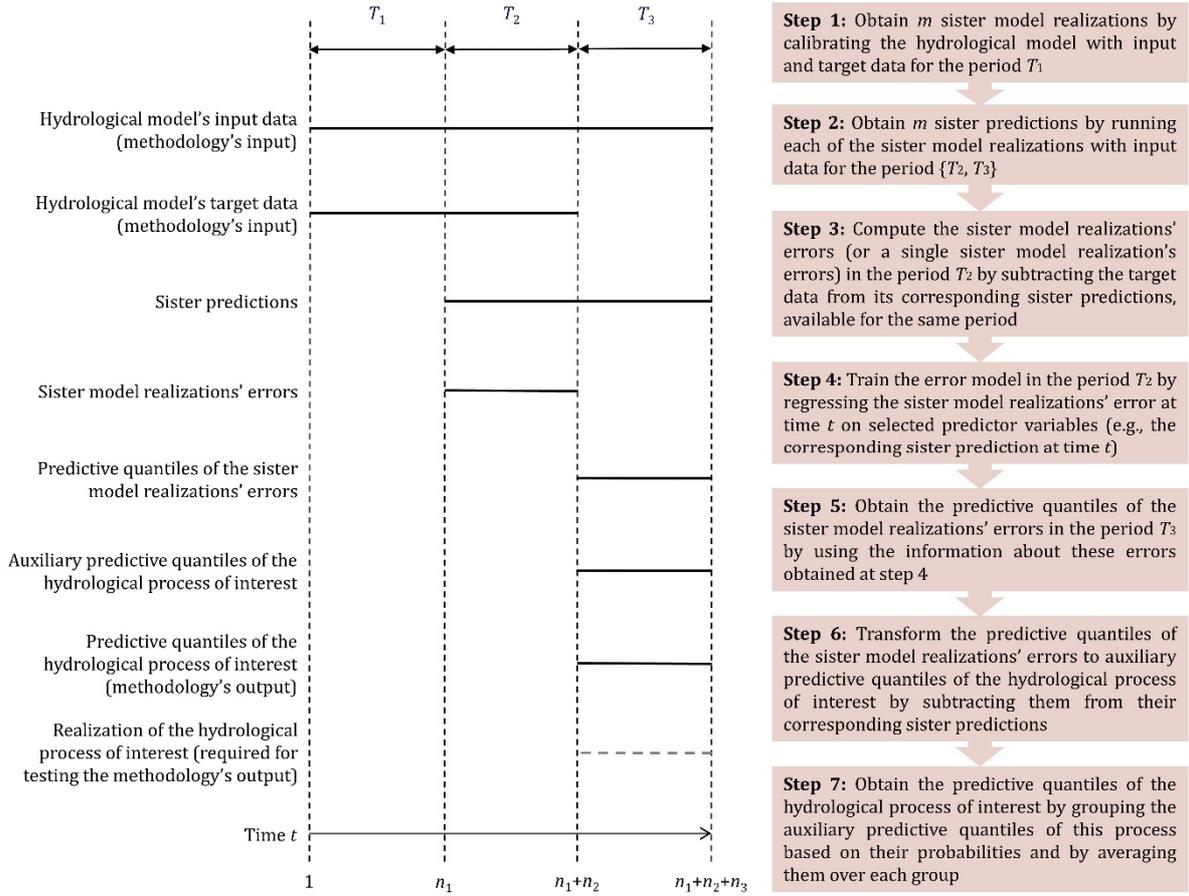

Figure 1. Schematic summarizing the methodology of the study. The sister model realizations are defined as variants of a single hydrological model, each using different parameter values. The latter are herein drawn from the respective simulated posterior distribution of model parameters, while they could be also obtained by using informal calibration schemes. Each sister model realization is used for obtaining a single point prediction, referred to as "sister prediction". The number of sister model realizations $m$ should be adequately large. The realization of the hydrological process of interest, considered unknown at the time of the prediction, is denoted with a light grey dashed line.

Let $\underline{y}$ be a stochastic process (typically a hydrological process, e.g., a streamflow or river discharge process), which is expressed in discrete time by Equation (1). In the following notations, the subscript of the variables $\underline{y}$ indicates the time $t$ or the time period. We wish to probabilistically predict the stochastic process $\underline{y}_{T_3}$ (hereafter referred to as "hydrological process of interest"), the realization of which is considered unknown at the time of the prediction. At this end, we assume the stochastic processes $\underline{x}_i$, where $i \in \{1, ..., n_0\}$ (denoting the sequential number assigned to each), and $\underline{x}$, which are informative about $\underline{y}$, and are expressed in discrete time by Equations (2) and (3) respectively. In the following notations the subscript of the variables $\underline{x}$ and second subscript of the variables $\underline{x}_i$ (separated by a comma from the first subscript) indicate the time $t$ or the time period. Let us also assume that the observations $\underline{x}_{T_3}$ are known at the time of the prediction.

$$\underline{y} := \underline{y}_T := (\underline{y}_1, ..., \underline{y}_{n_1}, \underline{y}_{(n_1+1)}, ..., \underline{y}_{(n_1+n_2)}, \underline{y}_{(n_1+n_2+1)}, ..., \underline{y}_{(n_1+n_2+n_3)})^T : (n_1+n_2+n_3) \times 1 \quad (1)$$

$$\underline{x}_i := \underline{x}_{i,T} := (\underline{x}_{i,1}, ..., \underline{x}_{i,n_1}, \underline{x}_{i,(n_1+1)}, ..., \underline{x}_{i,(n_1+n_2)}, \underline{x}_{i,(n_1+n_2+1)}, ..., \underline{x}_{i,(n_1+n_2+n_3)})^T : (n_1+n_2+n_3) \times 1 \quad (2)$$

$$\underline{x} := (\underline{x}_1, ..., \underline{x}_i, ..., \underline{x}_{n_0}) : (n_1+n_2+n_3) \times n_0 \quad (3)$$

Let $S$ be an arbitrary hydrological model, typically a (deterministic) point prediction model (e.g., a process-based hydrological model) that is suitable for predicting a variable $\underline{y}_t$ given the observations $\underline{x}_t$. The equations of such models may involve a variety of parameters, inputs (e.g., precipitation and temperature data at given time steps) and state variables (e.g., soil moisture levels, water table levels, snow cover). State variables are internal variables that describe the state of the catchment during simulation and change as a result of the modelling process (Beven 2012, pp. 5, 67, 176). Note also that $S$ could be used for forecasting $\underline{y}_t$ given forecasts instead of observations (see Klemeš 1986). Under this modelling approach, $\underline{y}_t$ is the dependent or response variable and $\underline{x}_t$ are the predictor variables at time $t$ (as assumed here), both expressed in stochastic terms. Let also $\underline{\theta}$ represent in stochastic terms the parameters of $S$ defined by Equation (4).

$$\underline{\theta} := (\underline{\theta}_1, ..., \underline{\theta}_j, ..., \underline{\theta}_n) : 1 \times n \quad (4)$$

Moreover, let us define $m$ variants of $S$, each using different parameters {$\boldsymbol{\theta}_k$, $k = 1, ..., m$}, where $m$ is adequately large (as large as our computational resources permit). These variants are hereafter referred to as "sister model realizations". The parameters {$\boldsymbol{\theta}_k$, $k = 1, ..., m$} are obtained by exploiting information from the period $T_1$. This exploitation can take various forms, such as simulation of the posterior distribution of $\underline{\theta}$ (by using Bayesian methods) or artificial simulation of $\underline{\theta}$ by using some type of randomization applied to a "best parameter estimate". The latter could be obtained by optimizing an objective function of our preference. Random selection of the parameters {$\boldsymbol{\theta}_k$, $k = 1, ..., m$} could also be an option. Herein we follow the Bayesian approach.

Once the sister model realizations are defined, they are all applied in the period {$T_2$, $T_3$}. The resulted $m$ sister predictions also extend in the period {$T_2$, $T_3$}. Let $\zeta_{k,t}$ be the point prediction at time $t \in$ {$T_2$, $T_3$} provided by the sister model realization that is defined by $\boldsymbol{\theta}_k$ (hereafter referred to as "$k^{th}$ sister model realization"). This point prediction is hereafter referred to as "$k^{th}$ sister prediction" at time $t$ to be distinguished from the remaining $m-1$ sister predictions at time $t$. In this case, $\zeta_{k,t}$ is obtained under the single-value transformation expressed by Equation (5), where $\boldsymbol{x}_t$ are the inputs to the model and $\boldsymbol{s}_t$ the values of the state variables at time $t$. We should note here again that the assumptions expressed through Equation (5) may vary from model to model. The input to $S$ could also include information from preceding time steps (e.g., $\boldsymbol{x}_{t-1}$, $\boldsymbol{x}_{t-2}$, $\boldsymbol{x}_{t-3}$, …), while the toy hydrological models used herein do not involve state variables $\boldsymbol{s}_t$ in their equations.

$$\zeta_{k,t} = S(\boldsymbol{\theta}_k, \boldsymbol{x}_t, \boldsymbol{s}_t) \qquad (5)$$

At time $t \in \{T_2, T_3\}$ the $k^{\text{th}}$ sister prediction $\zeta_{k,t}$ deviates from its target observation $y_t$, as expressed by Equation (6). The deviation $\varepsilon_{k,t}$, ignored by convention in the output of any point prediction model, is hereafter referred to as "$k^{\text{th}}$ sister model realization's error" at time $t$ and can be assumed as a realization of a random variable $\underline{\varepsilon}_k$. Such realizations are assumed to be informative about the uncertainty of the predictand $y_t$ conditional upon the $k^{\text{th}}$ sister prediction. Under this view, the sister model realizations' errors in the period $T_2$, i.e., $\boldsymbol{\varepsilon}_{k,T_2} \; \forall \; k \in \{1, …, m\}$, computed using the sister predictions $\boldsymbol{\zeta}_{k,T_2} \; \forall \; k \in \{1, …, m\}$ alongside with their targeted observations $\boldsymbol{y}_{T_2}$ (available), consist historical information that can be exploited for quantifying the predictive uncertainty in the period $T_3$.

$$\varepsilon_{k,t} := \zeta_{k,t} - y_t \qquad (6)$$

The methodology of the study is subdivided into three alternative variants, which differ to each other only in the exploitation of this historical information. For variants 1 and 2, we subsequently compute $\boldsymbol{\varepsilon}_{k,T_2} \; \forall \; k \in \{1, …, m\}$, while for variant 3 we compute $\boldsymbol{\varepsilon}_{k_0,T_2}$ for a randomly selected sister prediction $\boldsymbol{\zeta}_{k_0,T_2}$ with $k_0 \in \{1, …, m\}$. The exploitation of the related information is made by using an error model $M$, which falls into the category of statistical learning regression models that are suitable for predicting quantiles (see e.g., the quantile regression model detailed in Section 3.1). Let $e_{p,k,t}$ be the prediction of the conditional quantile with probability $p$ of the $k^{\text{th}}$ sister model realization's error at time $t$, obtained by using a trained version of $M$. Under this modelling approach, $\varepsilon_{k,t}$ is assumed to depend on selected informative variable(s). For reasons of simplicity, $\zeta_{k,t}$ is the only predictor variable considered herein for all three variants. The latter differ in the training of $M$. Specifically:

o    Variant 1 trains $M$ separately for each sister model realization. The training is,

therefore, made $m$ times, each time on a different dataset formed by using a different sister prediction $\zeta_{k,T_2}$ and its corresponding errors $\varepsilon_{k,T_2}$, where $k \in \{1, ..., m\}$;

o Variant 2 trains $M$ collectively for all sister model realizations. The training is, therefore, made once on a single dataset formed by using all sister predictions $\{\zeta_{k,T_2}, k = 1, ..., m\}$ and their corresponding errors $\{\varepsilon_{k,T_2}, k = 1, ..., m\}$;

o Variant 3 also trains $M$ once; however, the training here is made for an arbitrary sister model realization, i.e., on a dataset formed by using a randomly selected sister prediction $\zeta_{k_0,T_2}$ and its corresponding errors $\varepsilon_{k_0,T_2}$, where $k_0 \in \{1, ..., m\}$, under the assumption that $\varepsilon_{k_0,T_2}$ are informative about $\underline{\varepsilon}_k$ in general.

In what follows, the presentation is made for a single central prediction interval $(1 - \alpha)$, where $\alpha \in (0, 1)$, while the generalization to obtaining multiple central prediction intervals is straightforward. Let also $z_{p,k,t}$ be the obtained quantile with probability $p \in \{\alpha/2, 1 - \alpha/2\}$ of a variable of interest $\underline{y}_t$ conditional upon $\zeta_{k,t}$, hereafter referred to as "$k^{\text{th}}$ predictive quantile with probability $p$" of a variable of interest. Moreover, let $v_{p,t}$ be the finally delivered quantile with probability $p$ of a variable of interest $\underline{y}_t$, hereafter referred to simply as "predictive quantile with probability $p$" of this variable.

For each sister prediction $\zeta_{k,T_3}$, where $k \in \{1, ..., m\}$, we (a) predict the quantiles of the sister model realization's errors $\{e_{p,k,T_3}, p = \alpha/2, 1 - \alpha/2\}$ by using the information obtained in the preceding step, and (b) transform these predictive quantiles to "auxiliary predictive quantiles" of the hydrological process of interest $\{z_{p,k,T_3}, p = \alpha/2, 1 - \alpha/2\}$ by subtracting them from their corresponding sister prediction $\zeta_{k,T_3}$. At step (a) each trained version of $M$ is applied to predict the error quantiles of its corresponding sister prediction for variant 1, while for variants 2 and 3 the same trained version of $M$ is applied to predict the error quantiles of all sister predictions. Finally, at each time $t \in T_3$ we group the auxiliary predictive quantiles of the hydrological process of interest based on their corresponding probability $p$ (e.g., probability 0.95) to average them over each group. The resulted time series are the delivered quantile predictions $\{v_{p,T_3}, p = \alpha/2, 1 - \alpha/2\}$.

For the sake of completeness, variants 1–3 are algorithmically formulated in Appendix A. We note that these variants reduce to the same method in the case that a single point prediction is generated, i.e., for *m* = 1. In this case, the methodology of the study would fall into the category of basic two-stage post-processing methodologies using statistical learning regression models for quantile prediction (see e.g., Dogulu et al. 2015; López López et al. 2014; Papacharalampous et al. 2019c).

### 2.3 Remarks on the methodology of the study

The following remarks on the methodology of the study are important:

- The methodology of the study relies on the use of error models that by construction quantify predictive uncertainty, i.e., the total uncertainty of the predictand (parameter uncertainty included). This is why these error models have been exploited within basic hydrological post-processing methodologies. For instance, see the large-sample investigations in Papacharalampous et al. (2019c).

- The use of numerous parameter sets for the hydrological model is, thus, not a condition for properly considering parameter uncertainty. This is why the hydrological model's parameters can be obtained through informal calibration schemes.

- Simple quantile averaging is a novel methodological step compared to the original blueprint by Montanari and Koutsoyiannis (2012), and its variants by Sikorska et al. (2015) and Quilty et al. (2019). It is introduced herein to allow the accommodation of statistical learning regression models that are suitable for predicting quantiles into the methodology.

- Simple quantile averaging does not harm predictive uncertainty quantification. In fact, it works in the same way as simple PDF averaging. The latter has been exploited in hydrological post-processing concepts, e.g., by Vrugt (2018; see also 2019). We should note here again that, according to Lichtendahl et al. (2013), simple quantile averaging is as useful as (or even more useful than) simple PDF averaging. In Vrugt (2018, 2019) various point hydrological predictions are obtained by using different hydrological models (under a multi-model approach) and not by using a single hydrological model, as it is the case in the methodology of the study. These point hydrological predictions are first converted to PDF hydrological predictions (via post-processing) and then combined via (simple) PDF averaging.

## 2.4 Differences from other two-stage post-processing methodologies

Since the methodology of the study can be regarded as a set of variants of the MK blueprint methodology, some key changes with respect to the precursor methods should be underlined. These are the following:

- The methodology of the study is formulated to work with given data, i.e., it does not explicitly consider input data uncertainty (stemming e.g., from measurement errors; under the assumption of error-free data). Note that input data uncertainty could be considered (in a similar way to the one adopted in the precursor methods) if enough information is available to characterize it.

- The error models adopted in the precursor variants, i.e., the meta-Gaussian bivariate distribution model used in simulation mode by Montanari and Koutsoyiannis (2012), and the kNN model used by Sikorska et al. (2015) and Quilty et al. (2019), are here replaced by a statistical learning regression model that is suitable for predicting quantiles.

- Alternative options for the modelling of the sister model realizations' errors are provided. Additionally to variant 3, which extracts this type of information from a single sister prediction (as made in the MK blueprint methodology), we also include variants 1 and 2. These variants extract information about the hydrological model's error from all sister predictions.

- Ensemble predictions (i.e., individual predictions to be combined within an ensemble learning methodology; instead of ensemble simulations, i.e., individual simulations collectively composing an ensemble) are obtained and ensemble prediction averaging is involved. In fact, the methodology of the study falls into the category of ensemble learning methods (see e.g., Hastie et al. 2009, chapter 16), while the original variant, and the variants by Sikorska et al. (2015) and Quilty et al. (2019) are ensemble simulation methods.

Some key differences from other two-stage post-processing methodologies are also summarized subsequently:

- In contrast to basic two-stage post-processing methodologies using flexible quantile regression models (see e.g., Dogulu et al. 2015; López López et al. 2014; Solomatine and Shrestha 2009; Wani et al. 2017; Papacharalampous et al. 2019c), the methodology of the study is an ensemble learning methodology, as it combines multiple predictions to offer improved predictive performance.

- o In contrast to multi-model ensemble learning post-processing methodologies, the methodology of the study utilizes a single hydrological model.
- o In contrast to ensemble learning post-processing methodologies using multiple error models (see e.g., Tyralis et al. 2019b for the first stacked generalization approach to hydrological post-processing, and Papacharalampous et al. 2019c for an equal-weight combiner of six error models), the methodology of the study utilizes a single error model.

## 3. Experimental methodology

Here we present the experimental methodology adopted for the conducted toy model investigation. Statistical software information is summarized in Appendix B.

### 3.1 Statistical learning background

We implement three statistical learning regression models. Following the suggestions by Abrahart et al. (2008), we here emphasize on reproducibility and not on exhaustive descriptions of these models. The first regression model is the linear regression model (see e.g., James et al. 2013; Hastie et al. 2009) whose errors are zero-mean Gaussian i.i.d. (James et al. 2013). Its assumptions might not be efficient for real-world hydrological modelling applications; however, it offers the advantage of being interpretable (Hastie et al. 2009, page 43). We use it as described in Section 3.2, particularly focusing on the violation of the homoscedasticity assumption and on how its replacement with more flexible models under a predictive modelling view could result in improved probabilistic predictions. The second regression model is the quadratic regression model. The multiple linear regression model can accommodate quadratic (and polynomial) relationships, as described in James et al. (2013, chapter 3.3.2).

The quantile regression model by Koenker and Bassett (1978; see also Koenker 2005) is the third regression model implemented herein. This model focuses on modelling the conditional quantiles of the predictand variable by minimizing appropriate error functions (i.e., the quantile scores); therefore, it is appropriate for modelling heteroscedasticity (Koenker 2005, page 25), in contrast to the linear regression model. Related technical remarks can be found in Appendix C. Two advantages of the quantile regression model, as emphasized by López López et al.

(2014) are the robustness of the model with respect to outliers and the fact that no assumption is required for the PDF of the predictand variable. The quantile regression model offers a good compromise between interpretability (offered by the linear regression model) and flexibility (offered by more sophisticated statistical learning methods). Herein, it represents all regression models that can directly provide the predictive quantiles of the response variable, while they are also appropriate for modelling heteroscedasticity.

## 3.2 Toy modelling information

### 3.2.1 Toy data simulation

We simulate the three large toy datasets presented in Figure 2. Each of these datasets includes 12 000 pairs of ($x_t$, $y_t$) values, drawn i.i.d. from the populations described in Table 1. Benchmark remarks on the selection of the simulating models are also provided in Table 1.

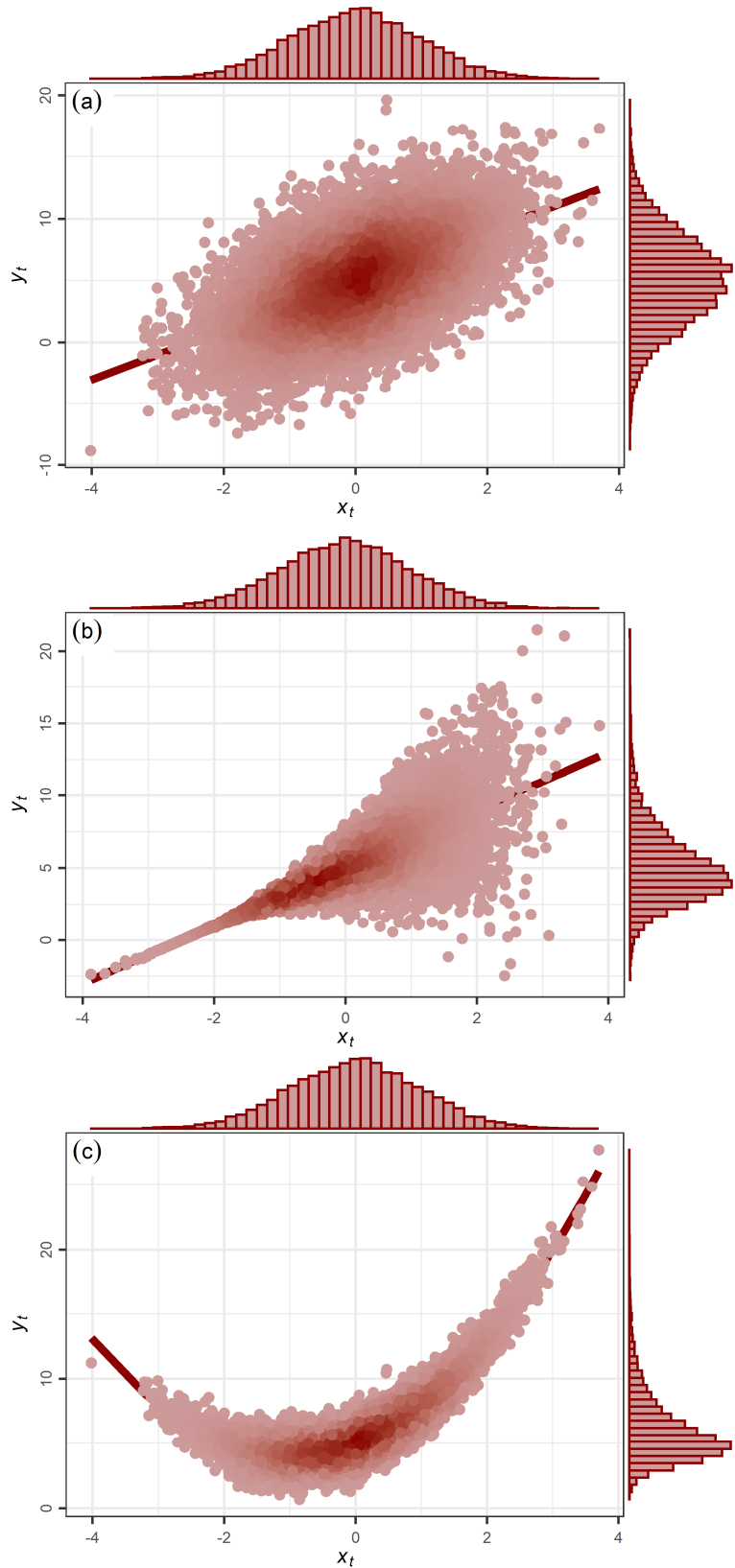

Figure 2. Toy datasets (a–c) 1–3. Details about their simulation are presented in Table 1. The pairs ($x_t$, $y_t$) are depicted with coloured bubbles (pink for low density and red for high density), while the red lines are the plots of the functions $y_t = f(x_t)$, i.e., the deterministic parts of the simulating models. The deviation in the vertical direction of a red line from any bubble is a realization of $\underline{u}_t$.

Table 1. Information about toy data simulation. The simulating models' types and parameters are selected to ensure a clear demonstration of the methodology of the study. The toy datasets are depicted in Figure 2. The function $f$ and the random variables $\underline{x}_t$, $\underline{u}_t$ and $\underline{y}_t$, where $t$ denotes the time, are defined as follows for each simulating model.

| Toy dataset | Simulating model | Remarks (see also Section 3.1) |
|---|---|---|
| 1 | $\underline{x}_t \sim N(\mu = 0, \sigma^2 = 1^2)$<br>$f(x) := 5 + 2x$<br>$\underline{u}_t \sim N(\mu = 0, \sigma^2 = 3^2)$<br>$\underline{y}_t := f(\underline{x}_t) + \underline{u}_t$ | i) There exists an analytical solution to delivering prediction intervals for this dataset, and ii) the assumptions of the simple linear regression model are proper for this dataset. |
| 2 | $\underline{x}_t \sim N(\mu = 0, \sigma^2 = 1^2)$<br>$f(x) := 5 + 2x$<br>$\underline{u}_t \sim N(\mu = 0, \sigma^2 = (0.2f(\underline{x}_t))^2)$<br>$\underline{y}_t := f(\underline{x}_t) + \underline{u}_t$ | The assumption of homoscedasticity of the error term (made by the simple linear regression model) is not proper for this dataset. |
| 3 | $\underline{x}_t \sim N(\mu = 0, \sigma^2 = 1^2)$<br>$f(x) := 5 + 2x + x^2$<br>$\underline{u}_t \sim N(\mu = 0, \sigma^2 = 1^2)$<br>$\underline{y}_t := f(\underline{x}_t) + \underline{u}_t$ | The assumption of linearity in the relationship between the predictor and the response (made by the simple linear regression model) is not proper for this dataset. |

### 3.2.2 Toy experiments, prediction schemes and expected outcomes

We conduct four toy experiments. Within each of these experiments we assess six ensemble schemes in obtaining interval predictions. The ensemble schemes are based on the methodology of the study, while they are defined by their underlying variants of this methodology and their adopted error models as prescribed by Table 2. They can be applied by using any point prediction model as (toy) hydrological model. Depending on the toy experiment, we adopt either the linear regression model or the quadratic regression model as toy hydrological models (see Table 3). These models are utilized by the ensemble schemes to generate point predictions of $y_t$ given $x_t$; therefore, $\underline{y}_t$ is the response variable and $\underline{x}_t$ is the predictor variable, both expressed in stochastic terms. A factor defining a toy experiment, together with the adopted toy hydrological model by all ensemble schemes, is the examined toy dataset (see Table 3). The toy datasets are presented in Section 3.2.1.

Table 2. Ensemble schemes assessed within the toy experiments of the study.

| Ensemble scheme | Variant of the methodology of the study | Outlined algorithm | Error model |
|---|---|---|---|
| 1 | 1 | Table A.1 | Linear regression model |
| 2 | 2 | Table A.2 | Linear regression model |
| 3 | 3 | Table A.3 | Linear regression model |
| 4 | 1 | Table A.1 | Quantile regression model |
| 5 | 2 | Table A.2 | Quantile regression model |
| 6 | 3 | Table A.3 | Quantile regression model |

Table 3. Toy experiments of the study. The toy datasets are presented in Section 3.2.1.

| Toy experiment | Toy dataset | Toy hydrological model for all ensemble schemes |
|---|---|---|
| 1 | 1 | Linear regression model |
| 2 | 2 | Linear regression model |
| 3 | 3 | Linear regression model |
| 4 | 3 | Quadratic regression model |

For the application of the ensemble schemes, detailed in Section 3.2.3, and following the definitions and notations provided in Section 2 (see also Appendix A), for each toy dataset we define the periods $T_1$ = {1, …, 1 000}, $T_2$ = {1 001, …, 2 000} and $T_3$ = {2 001, …, 12 000}. We include a large amount of information in the period $T_3$ to facilitate proper testing. To benchmark the toy results obtained using the methodology of the study, we also apply two basic probabilistic prediction schemes, namely the linear regression and quantile regression schemes. Their application is made according to Section 3.2.4. In particular for the case of toy experiment 1, we also consider the analytical solution provided by a Bayesian regression scheme, when the latter is applied under specific assumptions (see Section 3.2.4).

The only a priori theoretically expected outcomes in the conducted toy experiments are the following (see also Table 1; outcomes that need to be empirically proven are presented in Section 4):

o All three benchmark schemes are expected to perform well within toy experiment 1, in which the simple linear regression problem is solved for a large dataset. This problem is, in fact, the inverse problem with respect to the simulation of the therein utilized dataset for the linear regression model.

o The problem examined within toy experiment 2 is expected to be well-solved by the quantile regression model, while the solution provided by the linear regression model for the same problem is expected to be suboptimal. This problem could be viewed as an extension of the simple linear regression problem (James et al. 2013).

o The linear regression and quantile regression schemes are not the ideal models (when used with a single predictor) to be used for modelling a quadratic relationship. However, their predictions when both applied to toy dataset 3 are expected to not be equivalent.

o Ensemble schemes 1 and 4 are expected to provide the exact same solution with the basic post-processing methodologies using the linear regression and quantile regression models as error models respectively, when applied to toy datasets 1–3 with the simple linear regression model as toy hydrological model. The reason is theoretical; the problem solved by the error model for any point prediction provided by the simple linear regression model is practically the exact same one.

o This equivalence does not hold for any other toy hydrological model, e.g., the quadratic regression one. Therefore, within toy experiment 4 ensemble schemes 1 and 4 are expected to not be equivalent to basic two-stage post-processing methodologies.

### 3.2.3 Application of ensemble schemes

We describe the application of the ensemble schemes for one toy experiment, as all toy experiments are made in the same manner. The following steps are carried out once for all ensemble schemes:

1) We define 1 000 sister model realizations by obtaining the parameters {$\boldsymbol{\theta}_k$, $k$ = 1, …, 1 000} of the toy hydrological model. Specifically, we obtain 1 000 random samples of the joint posterior distribution of the toy hydrological model's parameters $\boldsymbol{\theta}$ and the variance of its error term $\sigma^2$ conditional on the observations of the period $T_1$. The joint posterior distribution is obtained by using a uniform prior distribution and an inverse prior distribution for $\boldsymbol{\theta}$ and $\sigma^2$ respectively, as detailed in Appendix C. Figure 3 summarizes information about the obtained {$\boldsymbol{\theta}_k$, $k$ = 1, …, 1 000} for toy experiment 1.

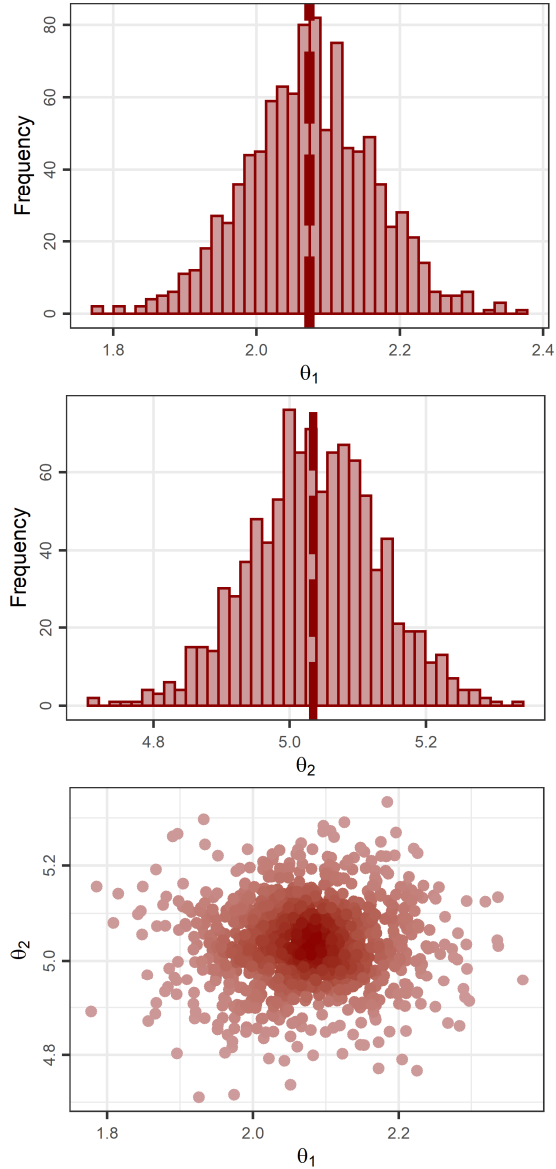

Figure 3. Simulated parameter values obtained using information from the period $T_1$ within toy experiment 1. The median $\theta_1$ and $\theta_2$ values are denoted with red thick dashed line on the presented histograms.

2) We obtain 1 000 sister predictions for the period $\{T_2, T_3\}$. Each sister prediction contains 11 000 values, while it is obtained by implementing a different sister model realization given the same information, i.e., input information for the period $\{T_2, T_3\}$.

3) By using the resulted sister predictions extending in the period $T_2$ alongside with their corresponding target values, we compute the sister model realizations' errors in the same period. The total number of the computed error values is 1 000 × 1 000 = 1 000 000. These values are considered informative about the sister model realization's errors in the period $T_3$ under the stationarity and ergodicity assumptions; therefore, they are used at the next step.

The following steps are carried out independently by each ensemble scheme:

4) We train the error model in the period $T_2$. Specifically, we regress the sister model realizations' error at time $t$ (response variable) on the sister prediction at time $t$ (predictor variable). The error model (linear regression or quantile regression), the number of the error model trainings (1 or 1 000) and the size of the training dataset(s) (1 000 000 or 1 000 pairs of values) depend on the ensemble scheme (see Section 2.2). We train the quantile regression model by using the algorithmic routine fully documented in Koenker and d'Orey (1987, 1994). Examples of training datasets are presented in Figure 4;

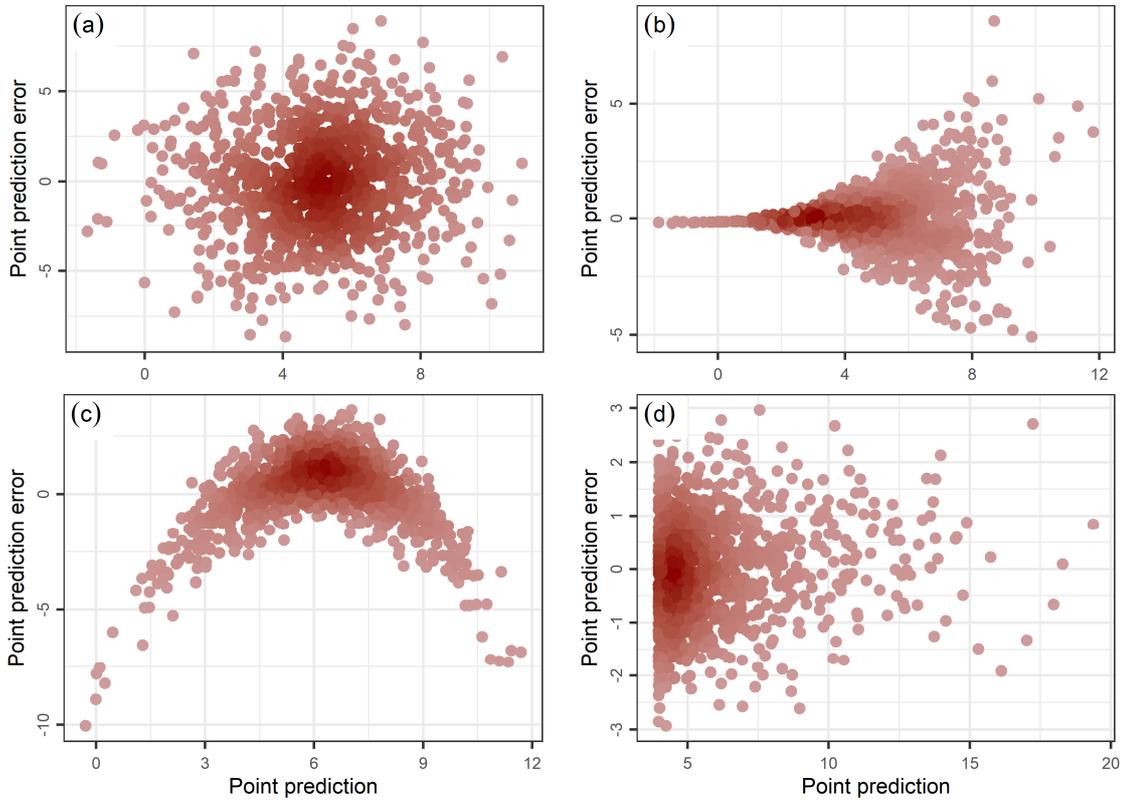

Figure 4. Error model training datasets for the ensemble schemes 3 and 6 within the toy experiments (a–d) 1–4.

5) We use each sister prediction extending in the period $T_3$ to predict a set of selected quantiles, specifically the quantiles with probability $p \in \{0.005, 0.0125, 0.025, 0.05, 0.10, 0.90, 0.95, 0.975, 0.9875, 0.995\}$, of its corresponding sister model realization's error. The predictions are made by exploiting information obtained in the preceding step (for details see Section 2.2). The result of this step is 1 000 probabilistic predictions for 10 000 data points, each consisting of 10 quantile predictions.

6) By subtracting each of these 1 000 × 10 = 10 000 quantile predictions from its

corresponding sister prediction, we obtain 1 000 auxiliary probabilistic predictions of the process of interest, each consisting of 10 quantile predictions.

7) Finally, we separately average, for each $p$ (as defined at point 5 above) and at each time $t \in T_3$, all the auxiliary predictive quantiles with probability $p$, i.e., 1 000 in number predictive quantiles, to obtain the finally delivered predictive quantile with probability $p$ at time $t$. The finally delivered predictive quantiles of the process of interest form the 99%, 97.5%, 95%, 90% and 80% central prediction intervals.

### 3.2.4 Application of benchmark schemes

The linear regression and quantile regression schemes are implemented by (a) training the linear regression and the quantile regression models respectively directly on the data from the period $\{T_1, T_2\}$ and, subsequently, by (b) applying the trained regression model in the period $T_3$ to predict the quantiles with probability $p$ (as defined at point 5 of Section 3.2.3) of the process of interest. The obtained predictive quantiles are then used to form the 99%, 97.5%, 95%, 90% and 80% central intervals. The predictor variable in regression is $\underline{x}_t$, expressed in stochastic terms.

The Bayesian regression scheme is trained by obtaining 1 000 random samples of the joint posterior distribution of the toy hydrological model's parameters $\underline{\boldsymbol{\theta}}$ and the variance of its error term $\underline{\sigma}^2$ conditional on the observations of the period $\{T_1, T_2\}$. A uniform prior distribution and an inverse prior distribution are used for $\underline{\boldsymbol{\theta}}$ and $\underline{\sigma}^2$ respectively, as detailed in Appendix C. Based on this joint posterior distribution of $\underline{\boldsymbol{\theta}}$ and $\underline{\sigma}^2$, the posterior predictive distribution for the period $T_3$ is obtained.

## 3.3 Performance assessment

In probabilistic modelling the aim is to maximize the sharpness of the predictive PDFs, subject to reliability (Gneiting and Katzfuss 2014). Reliability (or calibration) is the statistical correspondence between the probabilistic forecasts and the observations, while sharpness is the concentration of the predictive PDFs in absolute terms (Gneiting and Katzfuss 2014; see also Gneiting et al. 2007; Gneiting and Raftery 2007). We assess the reliability and sharpness of the obtained interval predictions by computing their coverage probabilities and average widths respectively. To simultaneously assess both these desired properties of the predictions, we also compute their average interval scores. The interval score by Winkler (1972), also known as Winkler score, rewards narrow prediction intervals, while penalizing prediction intervals missed by observations. The size of the penalty depends on the prediction interval (Gneiting and Raftery 2007). For a specific central prediction interval of level $(1 - \alpha)$, $0 < \alpha < 1$, extending in the period $T_3$, the coverage probability ($CP_\alpha$), average width ($AW_\alpha$) and average interval score ($AIS_\alpha$) are defined with Equations (7–9), where $I(\cdot)$ is the indicator function and $|T_3|$ is the number of the target data points included in the period $T_3$.

$$CP_\alpha := \sum_t I(y_t \in [v_{(\alpha/2),t}, v_{(1 - \alpha/2),t}])/|T_3| \qquad (7)$$

$$AW_\alpha := \sum_t (v_{(1 - \alpha/2),t} - v_{(\alpha/2),t})/|T_3| \qquad (8)$$

$$AIS_\alpha := \sum_t ((v_{(1 - \alpha/2),t} - v_{(\alpha/2),t}) + (2/\alpha)(v_{(\alpha/2),t} - y_t) I(y_t < v_{(\alpha/2),t}) + (2/\alpha)(y_t - v_{(1 - \alpha/2),t}) I(y_t > v_{(1 - \alpha/2),t}))/|T_3| \qquad (9)$$

Optimally, $CP_\alpha = (1 - \alpha)$, $\forall\ \alpha$ and for all datasets. For example, the perfect coverage probability for any 99% prediction interval is 0.99. On the contrary, the optimal average widths and average interval scores clearly depend on the examined dataset. For a specific dataset, the smaller the average width and average interval score of a specific prediction interval of level $(1 - \alpha)$ the better the interval prediction. Therefore, for benchmarking purposes we also compute the relative improvements ($RI_{\alpha,P_1,P_2}$), obtained when using a prediction interval $P_1$ of level $(1 - \alpha)$ (provided by a predictor of interest) with respect to another prediction interval $P_2$ of the same level (provided by a benchmark predictor) in terms of average interval score. This computation is made according to Equation (10). In this equation, $AIS_{\alpha,P_1}$ and $AIS_{\alpha,P_2}$ denote the average interval scores of the former and latter prediction intervals respectively.

$$RI_{\alpha,P_1,P_2} := (AIS_{\alpha,P_2} - AIS_{\alpha,P_1})/AIS_{\alpha,P_2} \qquad (10)$$

## 4. Experimental results, interpretations and illustrations

This section is devoted to the toy model investigation of the methodology of the study. This investigation is conducted within a purely statistical framework, while complementing Section 2 by largely facilitating the methodology's interpretation. The larger robustness in performance of this methodology compared to basic two-stage post-processing methodologies and its ability to harness the wisdom of the crowd are also illustrated using the obtained results.

### 4.1 Overall interpretation of the methodology of the study

In this section, we answer the following research questions (related to aims 2 and 4 of the study): (i) How does the methodology of the study and other two-stage hydrological post-processing methodologies work, and (ii) under which conditions these methodologies work well? In fact, although we focus on the methodology of the study, the presented toy examples can also be used to gain insight into two-stage hydrological post-processing in general. In Table 4, we present the coverage probabilities, average widths and average interval scores computed for the 99%, 97.5%, 95%, 90% and 80% prediction intervals obtained for all prediction schemes within toy experiment 1, while the respective results obtained for the toy experiments 2–4 are presented in Tables 5–7 respectively.

Table 4. Metric values computed for the prediction intervals delivered by the compared schemes for the period $T_3$ within the toy experiment 1.

| Metric | Prediction scheme | 99% prediction intervals | 97.5% prediction intervals | 95% prediction intervals | 90% prediction intervals | 80% prediction intervals |
|---|---|---|---|---|---|---|
| Coverage probability | Bayesian regression | 0.988 | 0.973 | 0.949 | 0.895 | 0.798 |
| | Linear regression | 0.989 | 0.973 | 0.948 | 0.897 | 0.798 |
| | Quantile regression | 0.986 | 0.971 | 0.945 | 0.891 | 0.802 |
| | Ensemble scheme 1 | 0.989 | 0.973 | 0.948 | 0.895 | 0.797 |
| | Ensemble scheme 2 | 0.989 | 0.972 | 0.947 | 0.895 | 0.797 |
| | Ensemble scheme 3 | 0.989 | 0.973 | 0.948 | 0.895 | 0.797 |
| | Ensemble scheme 4 | 0.987 | 0.967 | 0.951 | 0.890 | 0.805 |
| | Ensemble scheme 5 | 0.986 | 0.968 | 0.949 | 0.891 | 0.804 |
| | Ensemble scheme 6 | 0.987 | 0.967 | 0.951 | 0.890 | 0.805 |
| Average width | Bayesian regression | 15.29 | 13.35 | 11.69 | 9.82 | 7.65 |
| | Linear regression | 15.40 | 13.40 | 11.71 | 9.83 | 7.66 |
| | Quantile regression | 15.09 | 13.31 | 11.62 | 9.73 | 7.71 |
| | Ensemble scheme 1 | 15.36 | 13.36 | 11.68 | 9.80 | 7.63 |
| | Ensemble scheme 2 | 15.31 | 13.32 | 11.65 | 9.78 | 7.62 |
| | Ensemble scheme 3 | 15.36 | 13.36 | 11.68 | 9.80 | 7.63 |
| | Ensemble scheme 4 | 14.98 | 12.88 | 11.87 | 9.70 | 7.81 |
| | Ensemble scheme 5 | 14.94 | 13.03 | 11.81 | 9.73 | 7.76 |
| | Ensemble scheme 6 | 14.98 | 12.88 | 11.87 | 9.70 | 7.81 |
| Average interval score | Bayesian regression | 17.63 | 15.69 | 14.13 | 12.47 | 10.62 |
| | Linear regression | 17.47 | 15.67 | 14.11 | 12.46 | 10.61 |
| | Quantile regression | 17.77 | 15.81 | 14.23 | 12.52 | 10.61 |
| | Ensemble scheme 1 | 17.49 | 15.68 | 14.14 | 12.47 | 10.61 |
| | Ensemble scheme 2 | 17.49 | 15.69 | 14.14 | 12.47 | 10.61 |
| | Ensemble scheme 3 | 17.49 | 15.69 | 14.14 | 12.47 | 10.61 |
| | Ensemble scheme 4 | 17.56 | 15.82 | 14.18 | 12.52 | 10.65 |
| | Ensemble scheme 5 | 17.59 | 15.81 | 14.18 | 12.52 | 10.64 |
| | Ensemble scheme 6 | 17.57 | 15.82 | 14.18 | 12.52 | 10.65 |

Table 5. Metric values computed for the prediction intervals delivered by the compared schemes for the period $T_3$ within the toy experiment 2.

| Metric | Prediction scheme | 99% prediction intervals | 97.5% prediction intervals | 95% prediction intervals | 90% prediction intervals | 80% prediction intervals |
|---|---|---|---|---|---|---|
| Coverage probability | Linear regression | 0.975 | 0.962 | 0.946 | 0.919 | 0.864 |
| | Quantile regression | 0.994 | 0.986 | 0.967 | 0.918 | 0.824 |
| | Ensemble scheme 1 | 0.972 | 0.958 | 0.939 | 0.911 | 0.856 |
| | Ensemble scheme 2 | 0.972 | 0.958 | 0.939 | 0.911 | 0.856 |
| | Ensemble scheme 3 | 0.972 | 0.958 | 0.940 | 0.911 | 0.856 |
| | Ensemble scheme 4 | 0.994 | 0.982 | 0.960 | 0.905 | 0.819 |
| | Ensemble scheme 5 | 0.994 | 0.982 | 0.962 | 0.905 | 0.821 |
| | Ensemble scheme 6 | 0.994 | 0.982 | 0.961 | 0.906 | 0.819 |
| Average width | Linear regression | 7.74 | 6.73 | 5.89 | 4.94 | 3.84 |
| | Quantile regression | 7.75 | 6.45 | 5.21 | 3.99 | 2.92 |
| | Ensemble scheme 1 | 7.44 | 6.47 | 5.65 | 4.74 | 3.70 |
| | Ensemble scheme 2 | 7.41 | 6.45 | 5.64 | 4.73 | 3.69 |
| | Ensemble scheme 3 | 7.44 | 6.47 | 5.65 | 4.74 | 3.70 |
| | Ensemble scheme 4 | 7.61 | 6.06 | 4.93 | 3.75 | 2.86 |
| | Ensemble scheme 5 | 7.65 | 6.06 | 4.97 | 3.75 | 2.87 |
| | Ensemble scheme 6 | 7.63 | 6.08 | 4.94 | 3.77 | 2.87 |
| Average interval score | Linear regression | 14.08 | 10.50 | 8.54 | 6.90 | 5.40 |
| | Quantile regression | 8.86 | 7.56 | 6.37 | 5.31 | 4.31 |
| | Ensemble scheme 1 | 14.54 | 10.64 | 8.57 | 6.86 | 5.36 |
| | Ensemble scheme 2 | 14.60 | 10.66 | 8.57 | 6.87 | 5.36 |
| | Ensemble scheme 3 | 14.54 | 10.64 | 8.57 | 6.86 | 5.36 |
| | Ensemble scheme 4 | 8.86 | 7.48 | 6.33 | 5.33 | 4.31 |
| | Ensemble scheme 5 | 8.88 | 7.46 | 6.34 | 5.33 | 4.31 |
| | Ensemble scheme 6 | 8.88 | 7.49 | 6.33 | 5.33 | 4.31 |

Table 6. Metric values computed for the prediction intervals delivered by the compared schemes for the period $T_3$ within the toy experiment 3.

| Metric | Prediction scheme | 99% prediction intervals | 97.5% prediction intervals | 95% prediction intervals | 90% prediction intervals | 80% prediction intervals |
|---|---|---|---|---|---|---|
| Coverage probability | Linear regression | 0.979 | 0.970 | 0.959 | 0.933 | 0.865 |
| | Quantile regression | 0.989 | 0.975 | 0.950 | 0.909 | 0.813 |
| | Ensemble scheme 1 | 0.977 | 0.968 | 0.956 | 0.928 | 0.858 |
| | Ensemble scheme 2 | 0.977 | 0.968 | 0.956 | 0.927 | 0.857 |
| | Ensemble scheme 3 | 0.977 | 0.968 | 0.956 | 0.928 | 0.858 |
| | Ensemble scheme 4 | 0.990 | 0.977 | 0.945 | 0.903 | 0.802 |
| | Ensemble scheme 5 | 0.990 | 0.976 | 0.946 | 0.902 | 0.805 |
| | Ensemble scheme 6 | 0.990 | 0.976 | 0.945 | 0.902 | 0.801 |
| Average width | Linear regression | 9.04 | 7.87 | 6.88 | 5.77 | 4.50 |
| | Quantile regression | 10.88 | 8.73 | 6.93 | 5.53 | 3.99 |
| | Ensemble scheme 1 | 8.86 | 7.71 | 6.74 | 5.65 | 4.40 |
| | Ensemble scheme 2 | 8.84 | 7.69 | 6.72 | 5.64 | 4.40 |
| | Ensemble scheme 3 | 8.86 | 7.71 | 6.74 | 5.65 | 4.40 |
| | Ensemble scheme 4 | 10.83 | 8.57 | 6.48 | 5.37 | 3.87 |
| | Ensemble scheme 5 | 10.84 | 8.56 | 6.51 | 5.38 | 3.89 |
| | Ensemble scheme 6 | 10.83 | 8.57 | 6.49 | 5.37 | 3.87 |
| Average interval score | Linear regression | 16.13 | 11.90 | 9.60 | 7.72 | 6.10 |
| | Quantile regression | 12.73 | 10.47 | 8.90 | 7.45 | 5.98 |
| | Ensemble scheme 1 | 16.58 | 12.06 | 9.65 | 7.72 | 6.09 |
| | Ensemble scheme 2 | 16.68 | 12.09 | 9.66 | 7.73 | 6.09 |
| | Ensemble scheme 3 | 16.51 | 12.02 | 9.62 | 7.71 | 6.08 |
| | Ensemble scheme 4 | 12.46 | 10.56 | 8.98 | 7.44 | 5.98 |
| | Ensemble scheme 5 | 12.49 | 10.54 | 8.98 | 7.45 | 5.98 |
| | Ensemble scheme 6 | 12.48 | 10.57 | 8.99 | 7.45 | 5.99 |

Table 7. Metric values computed for the prediction intervals delivered by the compared schemes for the period $T_3$ within the toy experiment 4. The results of the linear regression and quantile regression schemes are repeated with respect to Table 6 for consistency in the presentation.

| Metric | Prediction scheme | 99% prediction intervals | 97.5% prediction intervals | 95% prediction intervals | 90% prediction intervals | 80% prediction intervals |
|---|---|---|---|---|---|---|
| Coverage probability | Linear regression | 0.979 | 0.970 | 0.959 | 0.933 | 0.865 |
| | Quantile regression | 0.989 | 0.975 | 0.950 | 0.909 | 0.813 |
| | Ensemble scheme 1 | 0.989 | 0.973 | 0.947 | 0.895 | 0.798 |
| | Ensemble scheme 2 | 0.989 | 0.972 | 0.946 | 0.894 | 0.798 |
| | Ensemble scheme 3 | 0.990 | 0.972 | 0.947 | 0.893 | 0.798 |
| | Ensemble scheme 4 | 0.986 | 0.968 | 0.949 | 0.893 | 0.802 |
| | Ensemble scheme 5 | 0.987 | 0.969 | 0.949 | 0.894 | 0.801 |
| | Ensemble scheme 6 | 0.987 | 0.968 | 0.950 | 0.892 | 0.803 |
| Average width | Linear regression | 9.04 | 7.87 | 6.88 | 5.77 | 4.50 |
| | Quantile regression | 10.88 | 8.73 | 6.93 | 5.53 | 3.99 |
| | Ensemble scheme 1 | 5.12 | 4.46 | 3.90 | 3.27 | 2.55 |
| | Ensemble scheme 2 | 5.11 | 4.44 | 3.89 | 3.26 | 2.54 |
| | Ensemble scheme 3 | 5.12 | 4.46 | 3.90 | 3.27 | 2.55 |
| | Ensemble scheme 4 | 5.00 | 4.34 | 3.93 | 3.25 | 2.57 |
| | Ensemble scheme 5 | 4.99 | 4.35 | 3.93 | 3.26 | 2.57 |
| | Ensemble scheme 6 | 5.01 | 4.32 | 3.95 | 3.24 | 2.59 |
| Average interval score | Linear regression | 16.13 | 11.90 | 9.60 | 7.72 | 6.10 |
| | Quantile regression | 12.73 | 10.47 | 8.90 | 7.45 | 5.98 |
| | Ensemble scheme 1 | 5.83 | 5.23 | 4.72 | 4.16 | 3.54 |
| | Ensemble scheme 2 | 5.83 | 5.23 | 4.72 | 4.16 | 3.54 |
| | Ensemble scheme 3 | 5.84 | 5.23 | 4.72 | 4.16 | 3.54 |
| | Ensemble scheme 4 | 5.86 | 5.27 | 4.72 | 4.16 | 3.54 |
| | Ensemble scheme 5 | 5.87 | 5.27 | 4.72 | 4.16 | 3.54 |
| | Ensemble scheme 6 | 5.86 | 5.27 | 4.72 | 4.16 | 3.54 |

Two considerations applying to each of toy experiments 1–3 (see Tables 4–6) are the following: (a) ensemble schemes 1–3, as well as ensemble schemes 4–6, are equivalent to each other on the examined normal data, and (b) each of the tested ensemble schemes is equivalent to its corresponding benchmark, i.e., ensemble schemes 1–3 and 4–6 perform as well as the linear regression and quantile regression schemes respectively. These two types of equivalence hold in terms of all three criteria examined. Consideration (a) also applies to the case of toy experiment 4 (see Table 4), while indicating that the three variants of the methodology of the study are equivalent in solving the examined problems. Moreover, consideration (b) can be viewed as an empirical proof that these problems are well-solved by the methodology of the study. The reason behind consideration (b) may become perceivable to some extent by comparing the original datasets (see Figure 2) with the datasets formed and used for training the incorporated quantile prediction models by the ensemble schemes (see Figure 4). Segments of the former datasets are used for training the benchmark schemes. In fact, the problems solved by each of the ensemble schemes and its corresponding benchmark seem to be of the same difficulty for toy experiments 1–3.

We have also tested the prediction schemes using shorter series (see e.g., the investigations of Appendix D and the large-sample experiment in Papacharalampous et al. 2019b). In that particular case for which the provided historical information is much less, the prediction schemes differentiate with each other in terms of performance. Nevertheless, by repeating the procedure an essentially large number of times with varying seed in the simulation of the datasets, we may observe long-run equivalence between specific prediction schemes, depending on the attributes of the datasets. For related discussions, the interested reader is referred to Appendix D.

One of the most important outcomes of the conducted toy model investigation is related to the satisfying coverage probabilities computed for all the ensemble schemes within toy experiment 1. Moreover, their good performance (equivalent to the performance of the Bayesian regression scheme and the two remaining benchmark schemes) in terms of average width of the prediction intervals and average interval score, observed within the same toy experiment, is important from an engineering point of view, as it points out that the methodology of the study does not lead to excessively precautionary design; see also the three criteria identified in Murphy (1993) for assessing the quality of predictions and the related discussions in Weijs et al. (2010), Ramos et al. (2010) and Papacharalampous et al. (2019a).

The performances of all prediction schemes differ for the toy implementations made on toy datasets 2 and 3, both in terms of coverage probability and average width; therefore, this differentiation is also manifested in the average interval scores. In fact, while both benchmark schemes are theoretically expected to be equally well-performing within toy experiment 1, quantile regression is theoretically expected to be better than linear regression within toy experiment 2, because of its advantage in modelling heteroscedasticity. We herein show that we can obtain equally good probabilistic predictions on normal data, by integrating the same model within the methodology of the study as error model. The interpretation of this outcome is straightforward; the incorporation of flexible models, such as the herein adopted quantile regression model may be the key to obtain efficient probabilistic predictions in specific modelling situations, including the hydrological modelling ones (see the comments on the violation of the homoscedasticity assumption in hydrological modelling, e.g., in Schoups and Vrugt 2010; Montanari and Koutsoyiannis 2012; Evin et al. 2013, 2014).

In greater detail, the numerical results of Table 5 can be summarized as follows. When using quantile regression instead of linear regression (within the methodology of the study) the average interval score is largely improved by around 40%, 30%, 35%, 30% and 25% for the 99%, 97.5%, 95%, 90% and 80% prediction intervals respectively. The respective relative improvements in terms of average width are around −3%, 6%, 13%, 22% and 22%, while the coverage probabilities computed for the predictions of the ensemble schemes 4–6 are essentially better than the coverage probabilities computed for the predictions of the ensemble schemes 1–3 for the 99%, 97.5%, 90% and 80% prediction intervals. The coverage probabilities of all ensemble schemes are comparable for the 95% prediction intervals. Typical performance differences observed within the study between probabilistic prediction schemes that use perfect and imperfect error models (with respect to modelling heteroscedasticity) are presented in Figure 5.

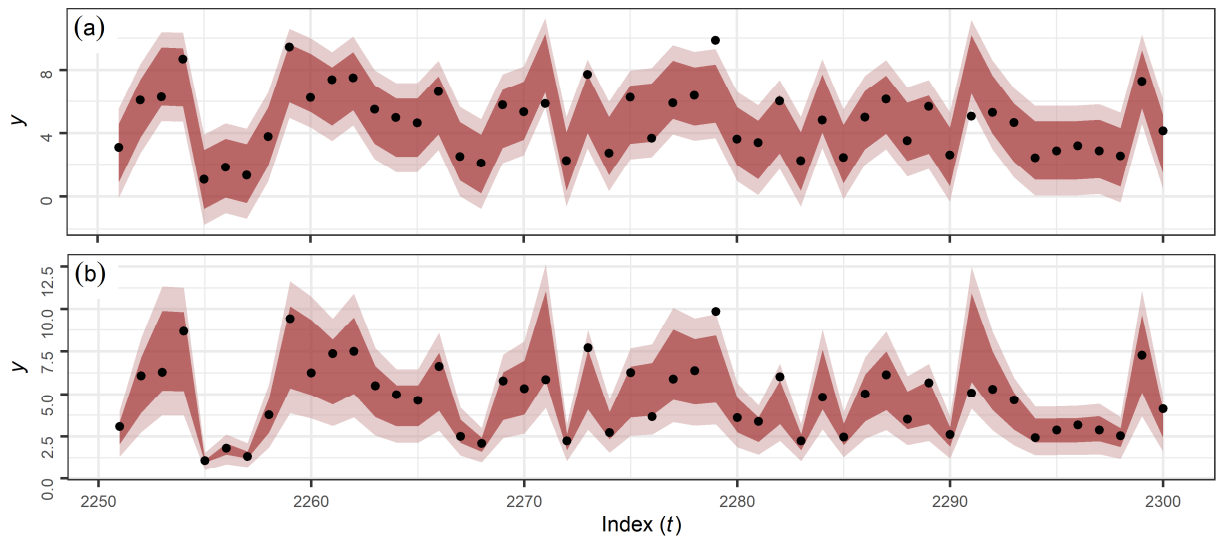

Figure 5. Toy solutions provided by ensemble schemes (a) 2 and (b) 5 within toy experiment 2 for a common 50-point sub-period of the period $T_3$. Black dots denote the targeted points, while light pink and dark pink ribbons denote the 95% and 80% prediction intervals respectively.

Moreover, within toy experiment 3 (see Table 6) we show that we can get probabilistic predictions with satisfactory coverage probabilities by using the methodology of the study, even when the incorporated toy hydrological model is imperfect (linear toy hydrological model for a quadratic relationship). This outcome is particularly important if we consider that process-based hydrological models are also imperfect. Specifically, we obtain perfect coverage probabilities by incorporating the quantile regression model within the methodology of the study, while the relative improvements in terms of average interval scores are around 25%, 13%, 8%, 4% and 2% for the 99%, 97.5%, 95%, 90% and 80% prediction intervals. The average widths, on the other hand, are better for ensemble schemes 4–6 only for the 95%, 90% and 80% prediction intervals, while they are much larger than those produced by ensemble schemes 1–3 for the 99% and 97.5% prediction intervals.

However, the average widths and average interval scores computed within toy experiment 3 for all ensemble schemes are found to be far from optimal, when contrasted to the results obtained within toy experiment 4 (see Table 7). The replacement of the imperfect (for toy dataset 3) toy hydrological model with a perfect one, has led to around 54%, 49%, 39%, 39% and 34% better average widths, and around 53%, 50%, 47%, 44% and 41% better average interval scores for the 99%, 97.5%, 95%, 90% and 80% prediction intervals respectively, as the latter are provided by ensemble schemes 4–6. In fact, the quality of the obtained probabilistic solution largely depends on the adopted toy hydrological model. A toy illustration of typical performance differences between probabilistic prediction schemes that use perfect and imperfect toy hydrological models is made in Figure 6.

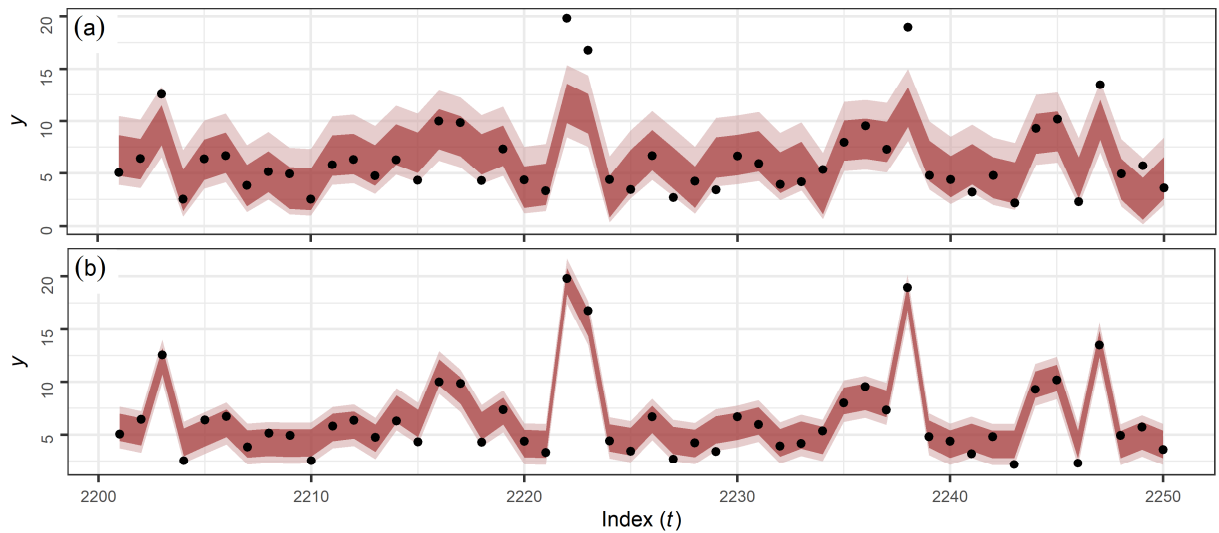

Figure 6. Toy solutions provided by ensemble scheme 5 within toy experiments (a) 3 and (b) 4 for a common 50-point sub-period of the period $T_3$. Black dots denote the targeted points, while light pink and dark pink ribbons denote the 95% and 80% prediction intervals respectively.

## 4.2 Improved robustness in hydrological post-processing

Here we present illustrative examples that can be used to gain further insight on how the methodology of the study works (aim 2 of the study) and to answer the following research question (related to aim 3 of the study): Why and when is it meaningful for someone to choose the methodology of the study over a basic two-stage post-processing methodology utilizing the same error model? In Figure 7, we present the relative improvements resulted within toy experiment 4 in terms of average interval score, when using the output of ensemble scheme 4, instead of each of the combined individual predictions, while in Figure 8 we present the respective relevant improvements provided by ensemble scheme 5. We observe that these relative improvements can be either positive or negative, while their mean is slightly higher than zero. Specifically, the average relative improvements computed for the histograms displayed in Figure 7 (Figure 8) are equal to 0.10%, 0.06%, 0.05%, 0.06% and 0.06% (0.20%, 0.10%, 0.13%, 0.14% and 0.12%) for the 99%, 97.5%, 95%, 90% and 80% prediction intervals respectively, with the results being analogous for the remaining ensemble schemes.

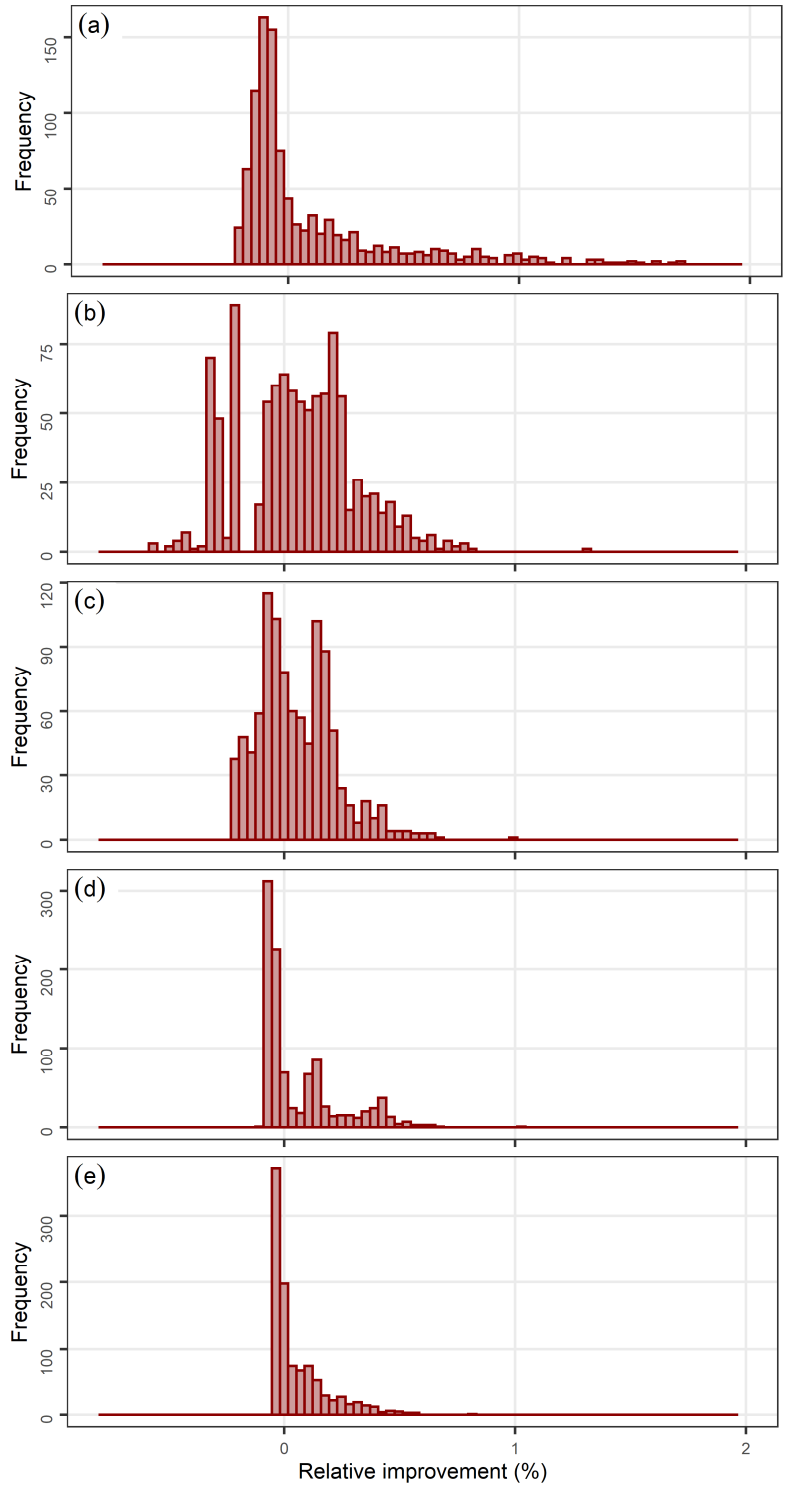

Figure 7. Relative improvements in terms of average interval score when using the output of ensemble scheme 4, i.e., the average of 1 000 probabilistic predictions, instead of each of the combined individual predictions. The relative improvements are computed for the (a) 99%, (b) 97.5%, (c) 95%, (d) 90% and (e) 80% prediction intervals obtained for the period $T_3$ within toy experiment 4. The horizontal axis has been truncated at −0.8% and 2%. Each histogram summarizes 1 000 values.

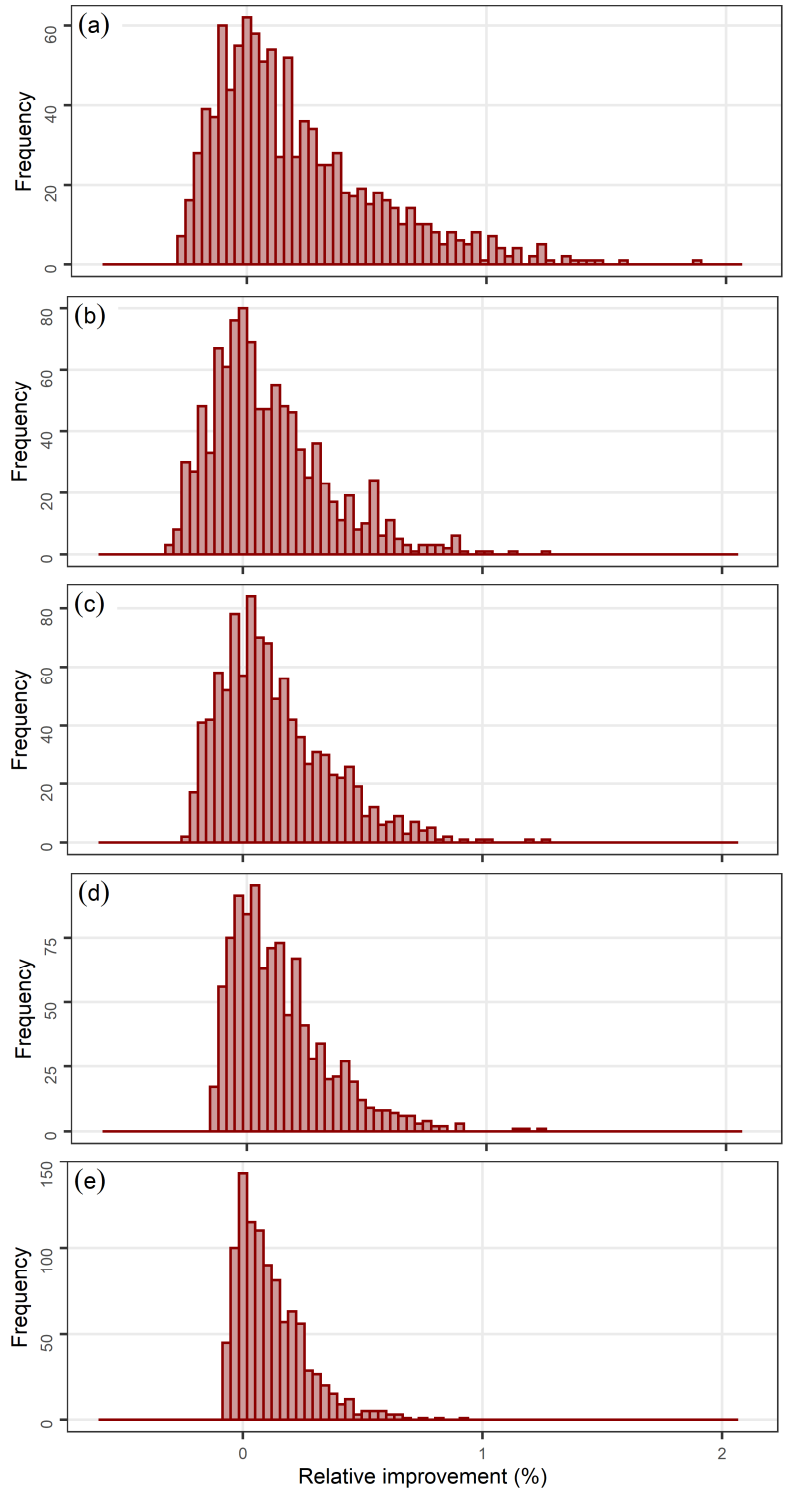

Figure 8. Relative improvements in terms of average interval score when using the output of ensemble scheme 5, i.e., the average of 1 000 probabilistic predictions, instead of each of the combined individual predictions. The relative improvements are computed for the (a) 99%, (b) 97.5%, (c) 95%, (d) 90% and (e) 80% prediction intervals obtained for the period $T_3$ within toy experiment 4. The horizontal axis has been truncated at −0.6% and 2%. Each histogram summarizes 1 000 values.

These average relative improvements computed for ensemble schemes 1 and 4 could be viewed as a direct comparison in terms of robustness in performance between the methodology of the study and a basic two-stage post-processing methodology (generating a single point hydrological prediction and, therefore, using a single set of hydrological model's parameters), the latter using the linear regression and quantile regression models respectively as error models. In fact, although many of the individual probabilistic predictions (obtained using different sister model realizations and, therefore, multiple sets of hydrological model's parameters) score better than the finally delivered one in the period $T_3$, we cannot know in advance which sister model realizations can be used for obtaining these better results and, therefore, should be preferred over the remaining ones within a basic post-processing methodology. By averaging numerous probabilistic predictions (obtained using the same number of different sister model realizations) we simply reduce the risk of delivering a probabilistic prediction of bad quality for the period $T_3$.

An important remark to be highlighted is that this risk can be high or low depending on the problem. In the toy problems examined herein the risk of delivering a probabilistic prediction of bad quality for the period $T_3$ (manifested in the magnitude of the relative improvements presented e.g., in Figures 7 and 8) is much lower than the respective risk that was found to be present in the rainfall-runoff problems examined by Papacharalampous et al. (2019b). In this latter study the computed relative improvements in terms of average interval score when using the output of the methodology of the study, instead of using each of the individual predictions combined for obtaining this output, range from about −330% to about 90%. (Negative relative improvements are computed for predictions that perform better that the prediction combination). Therefore, while it would not be that cost-efficient to use the methodology of the study for problems, such as the simple ones solved (for illustrative purposes) herein, it is cost-efficient from a risk management perspective to use this methodology (instead of a basic post-processing methodology) for probabilistic hydrological modelling applications.

Finally, for all ensemble schemes and all prediction intervals, the output of the methodology of the study is herein found to score slightly better than the average of the scores computed for each of the combined individual predictions in terms of average interval score. This latter information stands as an empirical proof that this methodology harnesses the wisdom of the crowd for the examined problem and in terms of average interval score. This useful property of the methodology of the study is further investigated by using rainfall-runoff datasets in Papacharalampous et al. (2019b).

## 5. Summary, discussion and conclusions

We have focused on the problem of probabilistically predicting hydrological variables, such as river discharge variables, by incorporating hydrological point prediction models, mainly falling into the category of deterministic process-based models, within stochastic modelling approaches. We have presented three novel variants of the blueprint methodology by Montanari and Koutsoyiannis (2012), also relying on the seminal work by Lichtendahl et al. (2013). In summary, the proposed methodology generates a large number of point predictions by utilizing a single hydrological model, yet with different parameter values. By solving a typical regression problem, these "sister predictions" are converted into auxiliary probabilistic predictions (consisted of quantile predictions), which are finally combined via simple quantile averaging. To the best of our knowledge, this is the first quantile averaging hydrological post-processing methodology that creates and exploits different information sets using a single model with different parameter values.

It is relevant to highlight that both the original blueprint and the herein introduced methodology fall into the family of two-stage probabilistic hydrological post-processing methodologies. Being mostly characterized by algorithmic-modelling-culture features (defined and analysed e.g., in Breiman 2001b and Shmueli 2010), and concomitant advantages and disadvantages, these methodologies aspire to achieve optimality in predictive performance in a fundamentally different way with respect to Bayesian joint inference methodologies for hydrological post-processing. Related information is provided in Papacharalampous et al. (2019b, Appendix A; see also Evin et al. 2014). In light of this information, the present study has been mostly devoted to finding modelling 'tricks' and concepts for maximizing predictive performance in two-stage hydrological post-processing by building on the original blueprint by Montanari and Koutsoyiannis (2012). An additional advantage offered by the latter with respect to other two-stage hydrological post-processing methodologies is its larger flexibility by perception, in the sense that it allows the formation and testing of various alternative configurations. This advantage is particularly important from a predictive modelling perspective.

A key improvement achieved herein compared to the original work by Montanari and Koutsoyiannis (2012), and the variants by Sikorska et al. (2015) and Quilty et al. (2019) in terms of flexibility in modelling is the use of statistical learning regression models that can directly provide predictive quantiles of the response variable, while they are also appropriate for modelling heteroscedasticity, such as the six machine learning algorithms examined in Papacharalampous et al. (2019c). These are quantile regression, generalized random forests for quantile regression, generalized random forests for quantile regression emulating quantile regression forests, gradient boosting machine, model-based boosting with linear models as base learners and quantile regression neural networks. Allowing the exploitation of the possibilities provided by this model category should, in fact, be regarded as a primary strength of the proposed methodology from a predictive modelling perspective.

Herein, we have demonstrated the usefulness of the proposed methodology and how our understanding of the system to be modelled can guide us to achieve better predictive modelling when using this methodology by conducting a toy model investigation. Within this investigation we have focused on the violation of the homoscedasticity assumption, when the latter is made in the modelling of the hydrological model's error, and on how the selection of an appropriate regression model for this task results in improved probabilistic predictions. We have also demonstrated the significance of using a better hydrological model for delivering probabilistic predictions that are simultaneously reliable and as sharp as possible. Finally, we have used the obtained toy results to show how the proposed methodology increases its robustness in performance by averaging many quantile predictions.

In spite of focusing on the introduced methodology, some of the obtained results can be used for gaining insight in general on how two-stage hydrological post-processing methodologies work and under which conditions their performance is maximized. The presented toy examples, demonstrating the key roles of both the statistical learning regression model and the hydrological model within a hydrological post-processing methodology, go beyond of some few exemplary (yet basic) toy tests that have already been made for the interpretation of methodologies for the quantification of the predictive hydrological uncertainty. Such tests mostly assume homoscedasticity and a perfect toy hydrological model, while here we are also inspired by recent simulation experiments that do not rely on these assumptions (see e.g., Evin et al. 2014; Renard et al. 2010; Vrugt et al. 2005).

The present paper is accompanied by Papacharalampous et al. (2019b), a study validating the herein introduced methodology and its key properties using a large amount of real-world data. Two simultaneously attractive and useful properties of this methodology that are extensively tested therein are its larger robustness in performance compared to the combined individual predictors and, by extension, compared to basic two-stage post-processing methodologies (which produce a single probabilistic prediction and, therefore, no prediction combination is made in their case), and its ability to "harness the wisdom of the crowd". The latter is defined in Lichtendahl et al. (2013, Section 5) as the property of some prediction combinations to score no worse –usually better– than the average score of the combined individual predictions. In fact, the larger the number of the combined quantile predictions (equal to the number of the generated sister predictions), the more robust the ensemble predictor and the more harnessed the wisdom of the crowd.

The proposed methodology is characterized by some additional strengths that are also particularly important from a predictive modelling point of view. First, it is computationally convenient in the sense that it can be easily expressed in algorithmic form (see Appendix A) and programmed using open source routines (see Appendix B; Papacharalampous et al. 2019b, Appendix B). Second, it offers certain modelling options that could be exploited to maximize predictive performance, as detailed in Papacharalampous et al. (2019b, Appendix F). For instance, variants 1 and 2 allow the exploitation by the error model of a large number of different information sets, instead of a single one (exploited by variant 3), thereby facilitating the enlargement of the sample space of the hydrological model's observed errors. This enlargement could be particularly important for modelling these errors using methods which do not extrapolate beyond the values of the training dataset, such as the quantile regression forests model (see the related theoretical information summarized by Tyralis et al. 2019a). Lastly, it allows the exploitation of the total amount of available information, in the sense that each sister prediction is herein converted into a probabilistic prediction (consisted of several quantile predictions) instead of a single simulation (randomly extracted from its predictive PDF; see the utilization of the meta-Gaussian bivariate distribution model in Montanari and Koutsoyiannis 2012; see also Kelly and Krzysztofowicz 1997).

Some limitations of the proposed methodology should also be considered. These include limitations implied by its two-stage nature (see Papacharalampous et al. 2019b, Appendix A), such as its shortcoming in terms of interpretability in modelling (especially in terms of producing interpretable parameter estimates) and its significant data length requirements (revealed e.g., in Appendix D). Although this latter limitation should be acknowledged herein and perhaps taken into consideration in real-world applications, (daily) datasets are usually essentially large. Moreover, in the companion paper by Papacharalampous et al. (2019b) it is empirically proven that, in practice, even when the available historical information is little, the proposed methodology is well-performing when implemented using the quantile regression model as error model.

Furthermore, the computational requirements of the proposed methodology are (at the moment) high when (i) computationally intensive procedures (e.g., Markov Chain Monte Carlo simulation sampling) are preferred for calibrating the hydrological model, and/or (ii) the error model is trained as implied by variant 1 or variant 2, unless the application is restricted to considering a small number of sister predictions. Note that a computationally convenient and simple algorithm is not necessarily computationally fast. It is also important to clarify that the above-outlined limitation holds only for applications to hundreds of catchments and timescales finer than the monthly one, and for implementations through regular personal computers. It does not hold for applications to a small number of catchments, and applications at the monthly and annual timescales. Still, large-scale applications at the daily timescale can be supported by variant 3, when this variant is implemented by using computationally fast algorithms for calibrating the hydrological model (see e.g., the calibration scheme tested in Papacharalampous et al. 2019b, Appendix E).

In addition to the above-discussed considerations and in contrast to several statistical methodologies for probabilistic prediction, such as the Bayesian methodology by Tyralis and Koutsoyiannis (2014), a well-known drawback of flexible statistical learning models for quantile prediction is their inappropriateness for modelling long-range dependence (see also Cox et al. 2018). Modelling this dependence when solving prediction problems is a frequently met concern in applied hydrology (see e.g., the large-scale investigations in Papacharalampous et al. 2018, 2019a). Nonetheless, empirical evidence (see e.g., Evin et al. 2014) suggests that the AR(1) assumption (in some sense allowed by the proposed methodology by using as a predictor variable in regression the hydrological model's prediction at time $t-1$) is adequate when modelling hydrological models' errors. In general, by including more than one predictor variables (e.g., the hydrological model's predictions at times $t$, $t$–1, $t$–2, etc.) in the regression settings we can increase the amount of the available information exploited and improve predictive performance, as it is empirically proven for rainfall-runoff modelling problems in Papacharalampous et al. (2019c).

Overall, the main trade-off to be considered when selecting between the methodology of the study and basic two-stage post-processing methodologies (utilizing the same error model) is the one between (a) the increased robustness in performance and the ability to harness the wisdom of the crowd, both offered by the former methodology, and (b) the significantly less computational requirements of a basic post-processing methodology. We believe that from a risk management standpoint this trade-off is worthy, as the large-sample experiment by Papacharalampous et al. (2019b) suggests.

## Appendix A   Algorithmic formulation of the methodology of the study

In this appendix, we algorithmically formulate the methodology of the study. The latter is sub-divided into three variants that are presented in detail in Tables A.1–A.3. The presentation is made for a single central prediction interval (1 – $\alpha$), where $\alpha \in (0, 1)$, formed by the predictive quantiles with probability $p$, where $p \in \{\alpha/2, 1 - \alpha/2\}$. The generalization to obtaining multiple central prediction intervals is straightforward.

Table A.1. Algorithmic formulation of the methodology of the study (variant 1). The presentation is made for a single central prediction interval (1 − $\alpha$), where $\alpha \in (0, 1)$, while the generalization to obtaining multiple central prediction intervals is straightforward. The repeated procedures are reported with different text alignment. Note that (i) the parameters $\{\boldsymbol{\theta}_k, k = 1, ..., m\}$ could be alternatively obtained through informal calibration schemes, and (ii) more predictors could be exploited in regression.

| Step | Procedure |
|---|---|
| 1 | Simulate the posterior distribution of $\underline{\boldsymbol{\theta}}$ using information for the time period $T_1$, i.e., obtain $\{\boldsymbol{\theta}_k, k = 1, ..., m\}$, for $m$ sufficiently large |
|  | **Repeat steps 2–6 $\forall\ k \in \{1, ..., m\}$** |
| 2 |     Obtain the $k^{\text{th}}$ sister prediction for the time period $\{T_2, T_3\}$, i.e., obtain $\boldsymbol{\zeta}_{k,\{T_2, T_3\}}$ according to: $\boldsymbol{\zeta}_{k,\{T_2, T_3\}} = S(\boldsymbol{\theta}_k, \boldsymbol{x}_{\{T_2, T_3\}})$ |
| 3 |     Compute the $k^{\text{th}}$ sister model realization's error for the time period $T_2$, i.e., obtain $\boldsymbol{\varepsilon}_{k,T_2}$ according to: $\boldsymbol{\varepsilon}_{k,T_2} = \boldsymbol{\zeta}_{k,T_2} - \boldsymbol{y}_{T_2}$ |
| 4 |     Regress the $k^{\text{th}}$ sister model realization's error $\varepsilon_{k,t}$ on the $k^{\text{th}}$ sister prediction $\zeta_{k,t}$ for the time period $T_2$, i.e., train $M$ between $\boldsymbol{\varepsilon}_{k,T_2}$ and $\boldsymbol{\zeta}_{k,T_2}$ |
| 5 |     Obtain the predictive quantiles of the $k^{\text{th}}$ sister model realization's error for the time period $T_3$ using the trained $M$, i.e., obtain $\boldsymbol{e}_{p,k,T_3}, \forall\ p \in \{\alpha/2, 1 - \alpha/2\}$, according to: $\boldsymbol{e}_{p,k,T_3} = M(\boldsymbol{\zeta}_{k,T_3})$ |
| 6 |     Obtain the $k^{\text{th}}$ predictive quantiles of the process of interest, i.e., obtain $\boldsymbol{z}_{p,k,T_3}, \forall\ p \in \{\alpha/2, 1 - \alpha/2\}$, according to:<br>○ $\boldsymbol{z}_{(\alpha/2),k,T_3} = \boldsymbol{\zeta}_{k,T_3} - \boldsymbol{e}_{(1-\alpha/2),k,T_3}$<br>○ $\boldsymbol{z}_{(1-\alpha/2),k,T_3} = \boldsymbol{\zeta}_{k,T_3} - \boldsymbol{e}_{(\alpha/2),k,T_3}$ |
| 7 | Obtain the predictive quantiles of the process of interest, i.e., obtain $\boldsymbol{v}_{p,T_3}, \forall\ p \in \{\alpha/2, 1 - \alpha/2\}$, by averaging separately $\forall\ t \in T_3$ the predictive quantiles $\{z_{p,k,t}, k = 1, ..., m\}$ according to: $v_{p,t} = \sum_{k=1}^{m} z_{p,k,t}$ |

Table A.2. Algorithmic formulation of the methodology of the study (variant 2). The presentation is made for a single central prediction interval (1 − $\alpha$), where $\alpha \in (0, 1)$, while the generalization to obtaining multiple central prediction intervals is straightforward. The repeated procedures are reported with different text alignment. Note that (i) the parameters $\{\boldsymbol{\theta}_k, k = 1, ..., m\}$ could be alternatively obtained through informal calibration schemes, and (ii) more predictors could be exploited in regression.

| Step | Procedure |
|---|---|
| 1 | Simulate the posterior distribution of $\underline{\boldsymbol{\theta}}$ using information for the time period $T_1$, i.e., obtain $\{\boldsymbol{\theta}_k, k = 1, ..., m\}$, for $m$ sufficiently large |
|  | **Repeat steps 2–3 $\forall\ k \in \{1, ..., m\}$** |
| 2 |     Obtain the $k^{\text{th}}$ sister prediction for the time period $\{T_2, T_3\}$, i.e., obtain $\boldsymbol{\zeta}_{k,\{T_2, T_3\}}$ according to: $\boldsymbol{\zeta}_{k,\{T_2, T_3\}} = S(\boldsymbol{\theta}_k, \boldsymbol{x}_{\{T_2, T_3\}})$ |
| 3 |     Compute the $k^{\text{th}}$ prediction error for the time period $T_2$, i.e., obtain $\boldsymbol{\varepsilon}_{k,T_2}$ according to: $\boldsymbol{\varepsilon}_{k,T_2} = \boldsymbol{\zeta}_{k,T_2} - \boldsymbol{y}_{T_2}$ |
| 4 | Regress the $k^{\text{th}}$ sister model realization's error $\varepsilon_{k,t}$ on the $k^{\text{th}}$ sister prediction $\zeta_{k,t}$ for the time period $T_2$, i.e., training of $M$ between $\boldsymbol{\varepsilon}_{k,T_2}$ and $\boldsymbol{\zeta}_{k,T_2}$. The training is performed collectively for all $k \in \{1, ..., m\}$. |
|  | **Repeat steps 5–6 $\forall\ k \in \{1, ..., m\}$** |
| 5 |     Obtain the predictive quantiles of the $k^{\text{th}}$ sister model realization's error for the time period $T_3$ using the trained $M$, i.e., obtain $\boldsymbol{e}_{p,k,T_3}, \forall\ p \in \{\alpha/2, 1 - \alpha/2\}$, according to: $\boldsymbol{e}_{p,k,T_3} = M(\boldsymbol{\zeta}_{k,T_3})$ |
| 6 |     Obtain the $k^{\text{th}}$ predictive quantiles of the process of interest, i.e., obtain $\boldsymbol{z}_{p,k,T_3}, \forall\ p \in \{\alpha/2, 1 - \alpha/2\}$, according to:<br>○ $\boldsymbol{z}_{(\alpha/2),k,T_3} = \boldsymbol{\zeta}_{k,T_3} - \boldsymbol{e}_{(1-\alpha/2),k,T_3}$<br>○ $\boldsymbol{z}_{(1-\alpha/2),k,T_3} = \boldsymbol{\zeta}_{k,T_3} - \boldsymbol{e}_{(\alpha/2),k,T_3}$ |
| 7 | Obtain the predictive quantiles of the process of interest, i.e., obtain $\boldsymbol{v}_{p,T_3}, \forall\ p \in \{\alpha/2, 1 - \alpha/2\}$, by averaging separately $\forall\ t \in T_3$ the predictive quantiles $\{z_{p,k,t}, k = 1, ..., m\}$ according to: $v_{p,t} = \sum_{k=1}^{m} z_{p,k,t}$ |

Table A.3. Algorithmic formulation of the methodology of the study (variant 3). The presentation is made for a single central prediction interval (1 − $\alpha$), where $\alpha \in (0, 1)$, while the generalization to obtaining multiple central prediction intervals is straightforward. The repeated procedures are reported with different text alignment. Note that (i) the parameters {$\theta_k$, $k$ = 1, ..., $m$} could be alternatively obtained through informal calibration schemes, and (ii) more predictors could be exploited in regression.

| Step | Procedure |
|---|---|
| 1 | Simulate the posterior distribution of $\boldsymbol{\theta}$ using information for the time period $T_1$, i.e., obtain {$\boldsymbol{\theta}_k$, $k$ = 1, ..., $m$}, for $m$ sufficiently large |
|  | **Repeat step 2 $\forall$ $k \in$ {1, ..., $m$}** |
| 2 | Obtain the $k^{\text{th}}$ sister prediction for the time period {$T_2$, $T_3$}, i.e., obtain $\boldsymbol{\zeta}_{k,\{T_2, T_3\}}$ according to: $\boldsymbol{\zeta}_{k,\{T_2, T_3\}} = S(\boldsymbol{\theta}_k, \boldsymbol{x}_{\{T_2, T_3\}})$ |
| 3 | Select a random $k_o \in$ {1, ..., $m$} |
| 4 | Compute the $k_o^{\text{th}}$ sister model realization's error for the time period $T_2$, i.e., obtain $\boldsymbol{\varepsilon}_{k_o,T_2}$ according to: $\boldsymbol{\varepsilon}_{k_o,T_2} = \boldsymbol{\zeta}_{k_o,T_2} - \boldsymbol{y}_{T_2}$ |
| 5 | Regress the $k_o^{\text{th}}$ sister model realization's error $\varepsilon_{k_o,t}$ on the $k_o^{\text{th}}$ sister prediction $\zeta_{k_o,t}$ for the time period $T_2$, i.e., train $M$ between $\boldsymbol{\varepsilon}_{k_o,T_2}$ and $\boldsymbol{\zeta}_{k_o,T_2}$ |
|  | **Repeat steps 6–7 $\forall$ $k \in$ {1, ..., $m$}** |
| 6 | Obtain the predictive quantiles of the $k^{\text{th}}$ sister model realization's error for the time period $T_3$ using the trained $M$, i.e., obtain of $\boldsymbol{e}_{p,k,T_3}$, $\forall$ $p \in$ {$\alpha/2$, 1 − $\alpha/2$}, according to: $\boldsymbol{e}_{p,k,T_3} = M(\boldsymbol{\zeta}_{k,T_3})$ |
| 7 | Obtain the $k^{\text{th}}$ predictive quantiles of the process of interest, i.e., obtain $\boldsymbol{z}_{p,k,T_3}$, $\forall$ $p \in$ {$\alpha/2$, 1 − $\alpha/2$}, according to: <br> ○ $\boldsymbol{z}_{(\alpha/2),k,T_3} = \boldsymbol{\zeta}_{k,T_3} - \boldsymbol{e}_{(1-\alpha/2),k,T_3}$ <br> ○ $\boldsymbol{z}_{(1-\alpha/2),k,T_3} = \boldsymbol{\zeta}_{k,T_3} - \boldsymbol{e}_{(\alpha/2),k,T_3}$ |
| 8 | Obtain the predictive quantiles of the process of interest, i.e., obtain $\boldsymbol{v}_{p,T_3}$, $\forall$ $p \in$ {$\alpha/2$, 1 − $\alpha/2$}, by averaging separately $\forall$ $t \in T_3$ the predictive quantiles {$z_{p,k,t}$, $k$ = 1, ..., $m$} according to: $v_{p,t} = \sum_{k=1}^{m} z_{p,k,t}$ |

## Appendix B    Statistical software information

The analyses and visualizations have been performed in R Programming Language (R Core Team 2019). We have used the following contributed R packages: BayesSummaryStatLM (Savel'ev et al. 2015), data.table (Dowle and Srinivasan 2019), devtools (Wickham et al. 2019b), gdata (Warnes et al. 2017), ggExtra (Attali 2017), ggplot2 (Wickham 2016a; Wickham et al. 2019a), knitr (Xie 2014, 2015, 2019), MASS (Venables and Ripley 2002; Ripley 2019), matrixStats (Bengtsson 2018), plyr (Wickham 2011, 2016b), quantreg (Koenker 2019), reshape (Wickham 2007, 2018), rmarkdown (Allaire et al. 2019) and tidyr (Wickham and Henry 2019).

## Appendix C    Technical remarks on the adopted modelling procedures

*C.1    Technical remarks on parameter estimation for the toy hydrological models*

Some technical remarks on the simulation of the posterior distributions of the parameters of the toy hydrological models should be made. These remarks are a summary of the information provided by Savel'ev et al. (2015). They are made for the case of the simple linear regression model (adopted as toy hydrological model within toy experiments 1–3), while the generalization to the multiple linear regression model (adopted as toy hydrological model within toy experiment 4) is straightforward.

Let us assume the simple linear regression model, expressed by Equations (C.1) and (C.2). In these equations, $\underline{y}$ and $\underline{x}$ are the predictand and predictor variables respectively, $\theta_0$ and $\theta_1$ are the regression coefficients, and $\underline{\varepsilon}_o$ is the fixed-variance error term, assumed i.i.d and normal.

$$\underline{y} = \theta_0 + \theta_1 \underline{x} + \underline{\varepsilon}_o \tag{C.1}$$

$$\underline{\varepsilon}_o \sim N(\mu_o = 0, \sigma_o{}^2) \tag{C.2}$$

Let us also assume that we are given a historical sample $\{(x_i, y_i), i = 1, …, \beta\}$, which could be also expressed by Equations (C.3), (C.4) and (C.5).

$$\boldsymbol{x}_{\{1, …, \beta\}} := (x_1, …, x_\beta)^T : \beta \times 1 \tag{C.3}$$

$$\boldsymbol{x}_B := [(1, …, 1)^T, \boldsymbol{x}_{\{1, …, \beta\}}] = [(1, …, 1)^T, (x_1, …, x_\beta)^T] : \beta \times 2 \tag{C.4}$$

$$\boldsymbol{y}_B := \boldsymbol{y}_{\{1, …, \beta\}} := (y_1, …, y_\beta)^T : \beta \times 1 \tag{C.5}$$

This sample can be exploited for simulating the posterior joint distribution of $\underline{\theta}_0$, $\underline{\theta}_1$ and $\underline{\sigma}^2$ by using the herein adopted Gibbs sampler. The latter is described by Equations (C.6) and (C.7), where $N_2$ denotes the bivariate normal distribution, $\boldsymbol{x}_B{}'$ the transpose of $\boldsymbol{x}_M$, Inv-Gamma the inverse gamma distribution and $(\theta_0, \theta_1)'$ the transpose of $(\theta_0, \theta_1)$.

$$\underline{\theta}_0, \underline{\theta}_1 \mid \sigma^2, \boldsymbol{x}_B, \boldsymbol{y}_B \sim N_2((\boldsymbol{x}_B{}' \boldsymbol{x}_B)^{-1} (\boldsymbol{x}_B{}' \boldsymbol{y}_B), \sigma^2 (\boldsymbol{x}_B{}' \boldsymbol{x}_B)^{-1}) \tag{C.6}$$

$$\underline{\sigma}^2 \mid \theta_0, \theta_1, \boldsymbol{x}_B, \boldsymbol{y}_B \sim \text{Inv-Gamma}(M/2, (\boldsymbol{y}_B{}' \boldsymbol{y}_B - (\theta_0, \theta_1)' \boldsymbol{x}_B{}' \boldsymbol{y}_B - \boldsymbol{y}_B{}' \boldsymbol{x}_B (\theta_0, \theta_1) + (\theta_0, \theta_1)' \boldsymbol{x}_B{}' \boldsymbol{x}_B (\theta_0, \theta_1))^{-1}/2) \tag{C.7}$$

The entire posterior joint distribution of $\underline{\theta}_0$, $\underline{\theta}_1$ and $\underline{\sigma}^2$ is exploited for producing probabilistic predictions through the Bayesian regression scheme (benchmark within toy experiment 1) according to the definition of prediction intervals. On the contrary, only the simulated $\theta_0$ and $\theta_1$ values are exploited by the ensemble schemes, while the simulated $\sigma^2$ values are discarded. For the ensemble schemes, the error of the (toy) hydrological model is modelled at a subsequent stage by using a different and independent historical sample, according to the remarks provided in the next section.

## C.2 Technical remarks on the application of the error models

Some technical remarks on the application of the error models should also be made. These remarks focus, among others, on the appropriateness of the quantile regression model for modelling heteroscedasticity. They are made by compiling information that mostly originates from Neter et al. (1983), Koenker and Hallock (2001), Koenker (2017) and Waldmann (2018).

Let us assume that we are interested in modelling the relationship between the random variables $\underline{y}$ and $\underline{x}$ given a training sample $\{(x_j, y_j), j = 1, …, γ\}$, so that we can probabilistically predict $\underline{y}$ conditional on $x$ in general later on. Let also $y_p(x)$ denote a quantile with probability $p \in (0, …, 1)$ of $\underline{y}$ conditional on $x$.

In summary, the quantile regression model is trained on the given sample separately for each probability $p$ by:

o  Assuming that all quantiles with probability $p$ share a common linear relationship with $\underline{x}$ expressed by Equation (C.8), where $θ_{0,p}$ and $θ_{1,p}$ are the regression coefficients to be estimated.

$$y_p(x) = θ_{0,p} + θ_{1,p}\, x \tag{C.8}$$

o  Optimizing the objective expressed by Equations (C.9) and (C.10) to estimate $θ_{0,p}$ and $θ_{1,p}$. Note that the right side of Equation (C.9) has been obtained by also exploiting Equation (C.8) above.

$$u_j := y_{p,j}(x_j) - y_j = θ_{0,p} + θ_{1,p}\, x_j - y_j \tag{C.9}$$

$$\min \sum_{j=1}^{γ} \bigl(p - \mathrm{I}(u_j < 0)\bigr)\, u_j \tag{C.10}$$

Therefore, by using the quantile regression model we are able to model quantiles of random variables "independently of distributional assumptions yet conditional on the data" (Waldmann 2018), with the focus being on describing how selected quantiles of the response variable change with changes of the predictor variable(s). As a result, quantile regression is appropriate for modelling heteroscedasticity.

On the contrary, the linear regression model focuses on describing how the mean of the response variable changes with the changes of the predictor variables. For instance, for our simple linear regression problem assumed above, the linear model is trained on the given sample by:

o  Assuming a linear relationship for the mean $μ$ and fixed variance $σ^2$ for the residuals, as expressed by Equations (C.1) and (C.2). In Equation (C.1), $θ_0$ and $θ_1$ are the regression coefficients to be estimated during training.

o  Optimizing the objective expressed by Equation (C.11) to estimate $θ_0$ and $θ_1$.

$$\min \sum_{j=1}^{γ} ε_{0,j}^2 \tag{C.11}$$

With the estimation of $\theta_0$ and $\theta_1$ two degrees of freedom are lost; therefore, the mean square error MSE that is defined by Equation (C.12) could serve as unbiased estimator of $\sigma^2$ (Neter et al. 1983, p. 47).

$$\text{MSE} := \left(\sum_{j=1}^{\gamma} \varepsilon_{o,j}^2\right)/(\gamma - 2) \tag{C.12}$$

When $\gamma$ is large (in practice larger than 30), any new central prediction interval (1 – $\alpha$), where $\alpha \in (0, 1)$, can be approximated conditional on the new $x_j$ and the training sample exploited in a preceding step by using Equation (C.13), where $\Phi^{-1}$ is the inverse standard normal cumulative distribution function (Neter et al. 1983, p. 81).

$$(\theta_0 + \theta_1 x_j) \pm \Phi^{-1}(1 - \alpha/2)\,(\text{MSE})^{1/2} \tag{C.13}$$

An illustrative example of modelling heteroscedasticity by using the quantile regression model and a comparison with the solution provided by the linear regression model for the same problem are given in Figure C.1.

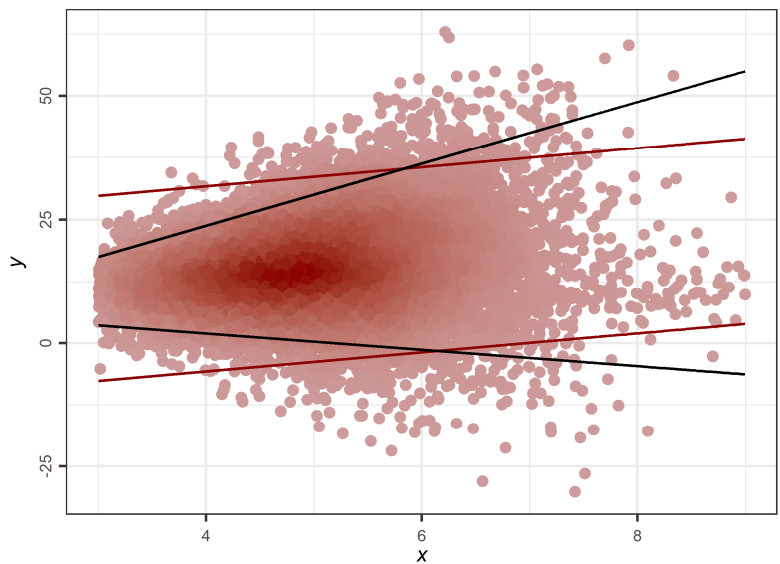

Figure C.1. Technical illustration of modelling heteroscedasticity using the quantile regression model and comparison with the linear regression model. The training data points are depicted with coloured bubbles (pink for low density and red for high density). The 90% central prediction intervals obtained for this training dataset using the linear regression and quantile regression models are depicted with red and black lines respectively.

## Appendix D    Additional investigations and derived interpretations

### D.1    Large-scale variant of toy experiment 1 using shorter toy datasets

Toy experiment 1 is particularly important because there exists an analytical solution to it, thereby allowing us to extensively explore under which conditions the data-driven solutions provided by the remaining schemes are adequate. This analytical solution is provided by the herein implemented Bayesian regression scheme. To further facilitate the interpretation of the proposed methodology (by answering questions related to aims 2 and 4 of the study), we here conduct the "type 1 additional investigations" of the study by repeating toy experiment 1 using shorter toy datasets. We run 500 repetitions, each time using a different toy dataset comprising 300 pairs of ($x_t$, $y_t$) values. These toy datasets result by following the same simulation procedure that was previously adopted for obtaining toy dataset 1 (see Table 1). Multiple runs are important in this case, because randomness can largely affect the results when relying on few data.

For each of the resulted toy datasets, we i) define the periods $T_1$ = {1, …, 100}, $T_2$ = {101, …, 200} and $T_3$ = {201, …, 300}, ii) run the three benchmark schemes according to Section 3.2.3, iii) run the six ensemble schemes according to Section 3.2.4 by adopting the linear regression model as toy hydrological model, and iv) compute the metric values for each delivered prediction according to Section 3.3. Finally, we compute the average metric values for each combination of prediction scheme and prediction interval. The coverage probability, average width and average interval score values are presented in Figures D.1, D.2 and D.3 respectively, while the average metric values are presented in Table D.1. Note that the here examined 500 toy datasets are all in the same scale; therefore, the average metric values are highly informative.

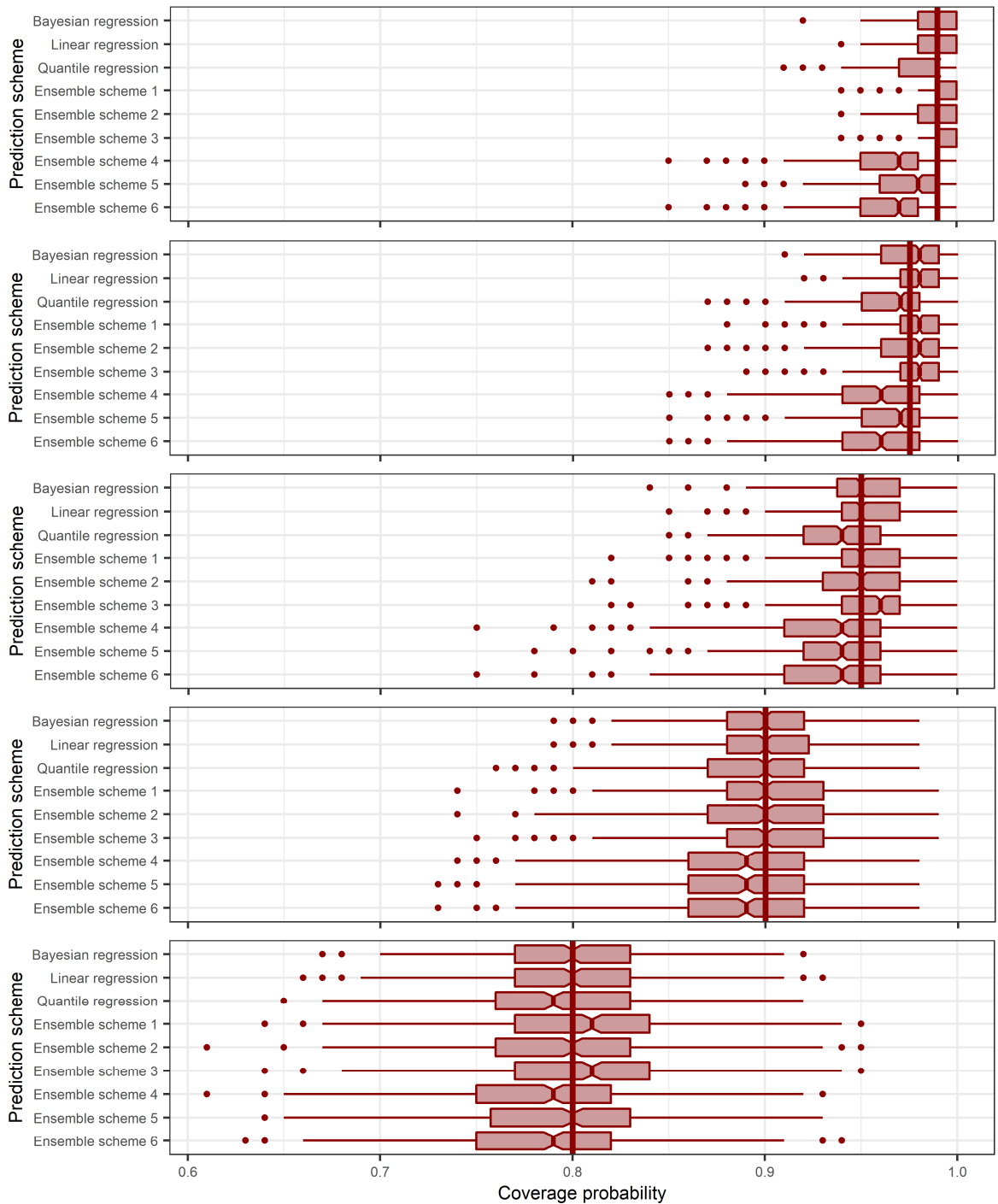

Figure D.1. Coverage probabilities computed for the 99%, 97.5%, 95%, 90% and 80% prediction intervals (from top to bottom) delivered by the compared schemes for the period $T_3$ within the type 1 additional investigations of the study. Each boxplot summarizes 500 values. The optimal values are denoted with red thick vertical lines.

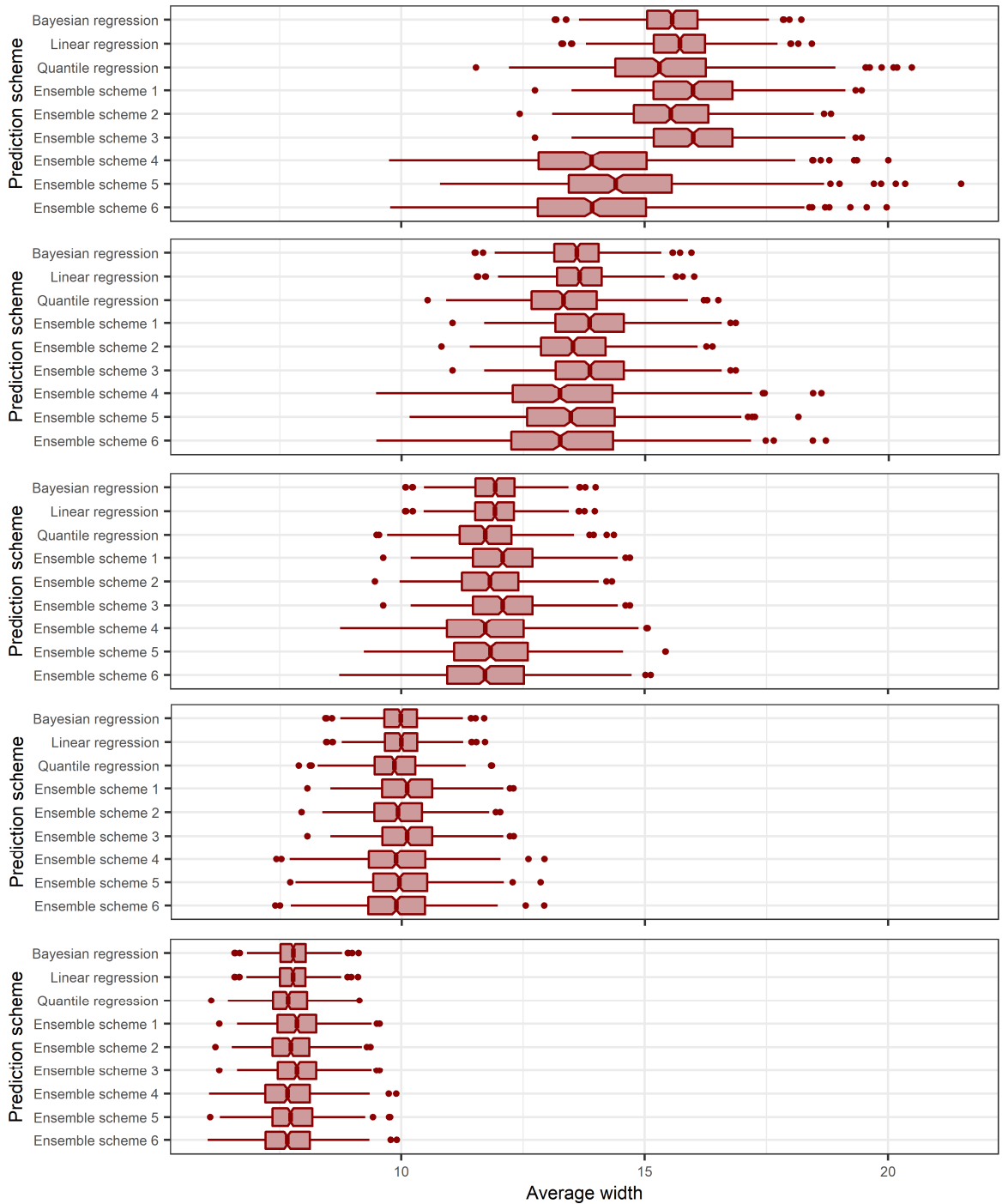

Figure D.2. Average widths computed for the 99%, 97.5%, 95%, 90% and 80% prediction intervals (from top to bottom) delivered by the compared schemes for the period $T_3$ within the type 1 additional investigations of the study. Each boxplot summarizes 500 values.

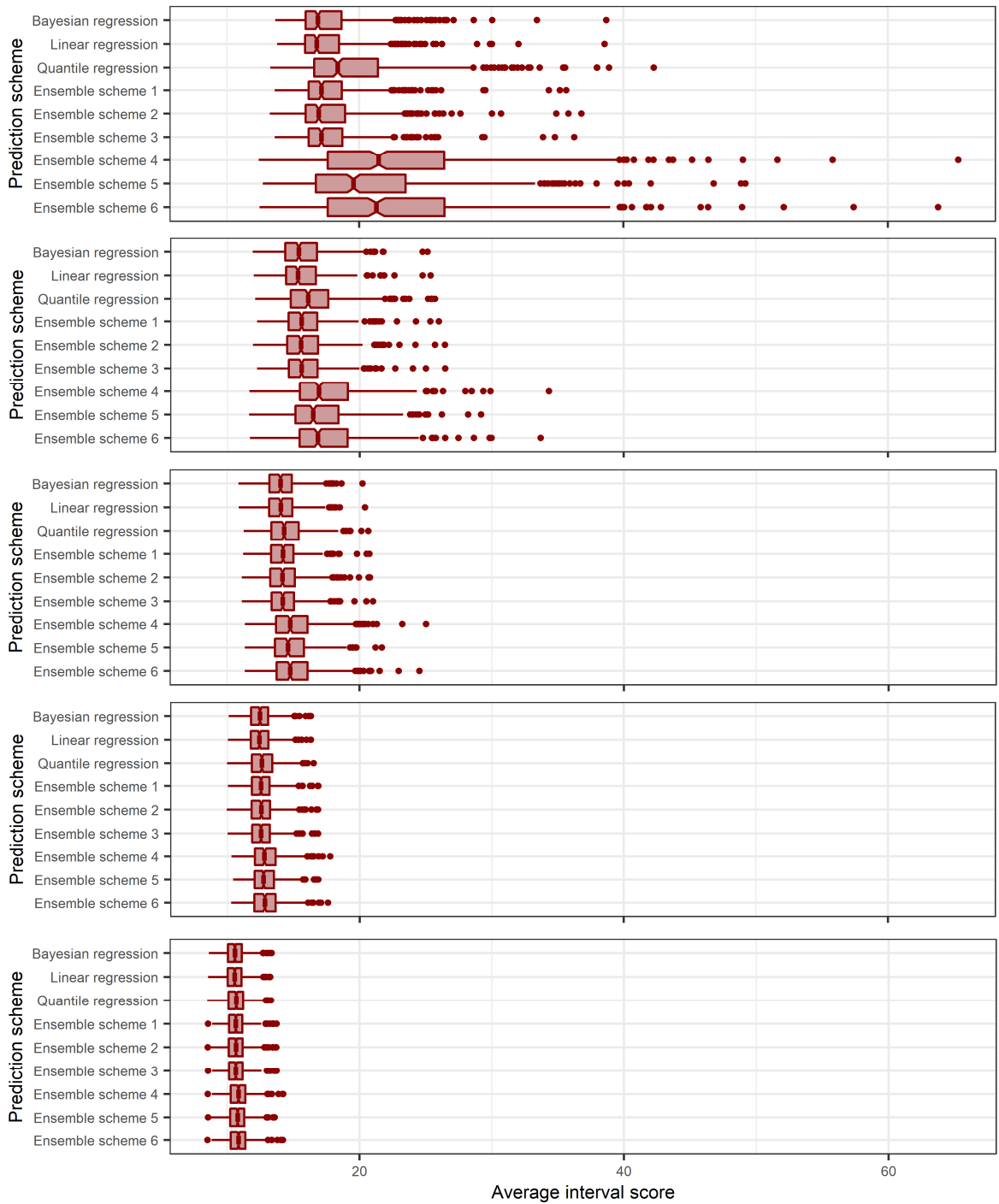

Figure D.3. Average interval scores computed for the 99%, 97.5%, 95%, 90% and 80% prediction intervals (from top to bottom) delivered by the compared schemes for the period $T_3$ within the type 1 additional investigations of the study. Each boxplot summarizes 500 values.

Table D.1. Average metric values computed for the prediction intervals delivered by the compared schemes for the period $T_3$ within the type 1 additional investigations of the study. Each presented value summarizes 500 metric values.

| Metric | Prediction scheme | 99% prediction intervals | 97.5% prediction intervals | 95% prediction intervals | 90% prediction intervals | 80% prediction intervals |
|---|---|---|---|---|---|---|
| Coverage probability | Bayesian regression | 0.989 | 0.975 | 0.949 | 0.899 | 0.801 |
| | Linear regression | 0.990 | 0.976 | 0.950 | 0.900 | 0.800 |
| | Quantile regression | 0.981 | 0.966 | 0.941 | 0.892 | 0.793 |
| | Ensemble scheme 1 | 0.991 | 0.976 | 0.952 | 0.902 | 0.803 |
| | Ensemble scheme 2 | 0.989 | 0.973 | 0.947 | 0.896 | 0.796 |
| | Ensemble scheme 3 | 0.991 | 0.976 | 0.952 | 0.902 | 0.803 |
| | Ensemble scheme 4 | 0.964 | 0.957 | 0.933 | 0.886 | 0.785 |
| | Ensemble scheme 5 | 0.973 | 0.962 | 0.939 | 0.891 | 0.792 |
| | Ensemble scheme 6 | 0.964 | 0.957 | 0.933 | 0.885 | 0.786 |
| Average width | Bayesian regression | 15.57 | 13.60 | 11.93 | 9.98 | 7.78 |
| | Linear regression | 15.72 | 13.65 | 11.92 | 9.99 | 7.77 |
| | Quantile regression | 15.43 | 13.36 | 11.75 | 9.86 | 7.70 |
| | Ensemble scheme 1 | 16.00 | 13.87 | 12.09 | 10.12 | 7.86 |
| | Ensemble scheme 2 | 15.54 | 13.53 | 11.83 | 9.93 | 7.73 |
| | Ensemble scheme 3 | 16.00 | 13.87 | 12.09 | 10.12 | 7.86 |
| | Ensemble scheme 4 | 13.98 | 13.32 | 11.71 | 9.90 | 7.67 |
| | Ensemble scheme 5 | 14.53 | 13.46 | 11.86 | 9.97 | 7.75 |
| | Ensemble scheme 6 | 13.99 | 13.32 | 11.71 | 9.90 | 7.67 |
| Average interval score | Bayesian regression | 17.72 | 15.72 | 14.18 | 12.51 | 10.62 |
| | Linear regression | 17.61 | 15.69 | 14.16 | 12.50 | 10.62 |
| | Quantile regression | 19.66 | 16.46 | 14.51 | 12.66 | 10.70 |
| | Ensemble scheme 1 | 17.88 | 15.91 | 14.31 | 12.60 | 10.69 |
| | Ensemble scheme 2 | 17.89 | 15.91 | 14.31 | 12.60 | 10.69 |
| | Ensemble scheme 3 | 17.88 | 15.91 | 14.31 | 12.60 | 10.69 |
| | Ensemble scheme 4 | 22.85 | 17.49 | 15.07 | 12.93 | 10.86 |
| | Ensemble scheme 5 | 20.97 | 16.97 | 14.83 | 12.83 | 10.80 |
| | Ensemble scheme 6 | 22.76 | 17.47 | 15.06 | 12.92 | 10.86 |

The main observations extracted from these investigations can be summarized as follows: (a) The linear regression scheme is equivalent to the Bayesian regression scheme in the long run, (b) ensemble schemes 1–3 perform almost as well as the two best-performing benchmarks (since their error modelling procedures are benefited from proper assumptions, i.e., prior knowledge on the system to be modelled), (c) ensemble schemes 4–6 are the worst-performing, (d) the quantile regression scheme exhibits a moderate performance, (e) ensemble schemes 1–3 are almost equivalent to each other, and (f) ensemble scheme 5 performs better than ensemble schemes 4 and 6. By comparing these observations with those extracted from toy experiment 1 (see Section 4.1), we understand that the quantile regression model needs to be "fed" with more data to reach its best performance, which in the case of the here examined data type is as good as the performance of the linear regression model. Note that this "data consuming" consideration stems from the statistical learning nature of the modelling process and, therefore, it applies to the linear regression model as well, yet to a smaller extent. It could also be viewed as a limitation of two-stage hydrological post-processing in general (see Papacharalampous et al. 2019b, Appendix A).

*D.2   Large-scale toy regression experiment with non-informative predictors*

We repeat the type 1 additional investigations of the study by using different toy datasets of the same number and length. The new datasets result from a simulating model that implies no dependence of **y** on **x**, specifically the following: $x_t \sim N(\mu = 0, \sigma^2 = 1^2)$ and $y_t \sim N(\mu = 0, \sigma^2 = 1^2)$. These investigations are hereafter referred to as "type 2 additional investigations". The analytical solution to the examined toy problem is provided by the "Bayesian non-regression" benchmark. This scheme assumes that $y_t \sim N(\mu, \sigma^2)$, prior independence of $\mu$ and $\sigma$, and a uniform prior distribution on $(\mu, \log\sigma)$. Then, the posterior distribution is a Student-$t$ distribution with location $\bar{y} = (1/n) \sum_{t=1}^{n} y_t$, scale $(1 + 1/n)^{0.5} ((1/(n-1)) \sum_{t=1}^{n}(y_t - \bar{y})^2)^{0.5}$, and $n - 1$ degrees of freedom, where $n$ is the number of data points included in the fitting sample (Gelman et al. 2004). In our case, the fitting sample is consisted of the first 200 data points of each simulated series, i.e., $n = 200$.

The main observations extracted from the type 2 additional investigations of the study can be summarized as follows (see Table D.2): (a) The Bayesian non-regression, Bayesian regression and linear regression benchmarks are equivalent in the long run, (b) ensemble schemes 1 and 2 perform almost as well as the three best-performing benchmarks, (c) ensemble schemes 3 and 6 are the worst-performing (mostly because of outliers), (d) ensemble scheme 5 and the quantile regression benchmark exhibit a moderate performance, and (e) ensemble scheme 4 exhibits better performance than ensemble scheme 6 and worse than ensemble scheme 5.

Table D.2. Average metric values computed for the prediction intervals delivered by the compared schemes for the period $T_3$ within the type 2 additional investigations of the study. Each presented value summarizes 500 metric values.

| Metric | Prediction scheme | 99% prediction intervals | 97.5% prediction intervals | 95% prediction intervals | 90% prediction intervals | 80% prediction intervals |
|---|---|---|---|---|---|---|
| Coverage probability | Bayesian non-regression | 0.990 | 0.976 | 0.950 | 0.901 | 0.801 |
| | Bayesian regression | 0.989 | 0.975 | 0.949 | 0.899 | 0.801 |
| | Linear regression | 0.990 | 0.976 | 0.950 | 0.900 | 0.800 |
| | Quantile regression | 0.981 | 0.966 | 0.941 | 0.892 | 0.793 |
| | Ensemble scheme 1 | 0.991 | 0.976 | 0.952 | 0.902 | 0.803 |
| | Ensemble scheme 2 | 0.988 | 0.972 | 0.947 | 0.895 | 0.795 |
| | Ensemble scheme 3 | 0.994 | 0.982 | 0.962 | 0.920 | 0.832 |
| | Ensemble scheme 4 | 0.964 | 0.957 | 0.933 | 0.886 | 0.785 |
| | Ensemble scheme 5 | 0.980 | 0.964 | 0.943 | 0.894 | 0.794 |
| | Ensemble scheme 6 | 0.893 | 0.895 | 0.877 | 0.835 | 0.736 |
| Average width | Bayesian non-regression | 5.23 | 4.54 | 3.96 | 3.32 | 2.58 |
| | Bayesian regression | 5.19 | 4.53 | 3.98 | 3.33 | 2.59 |
| | Linear regression | 5.24 | 4.55 | 3.97 | 3.33 | 2.59 |
| | Quantile regression | 5.14 | 4.45 | 3.92 | 3.29 | 2.57 |
| | Ensemble scheme 1 | 5.33 | 4.62 | 4.03 | 3.37 | 2.62 |
| | Ensemble scheme 2 | 5.14 | 4.48 | 3.91 | 3.28 | 2.56 |
| | Ensemble scheme 3 | 8.72 | 7.56 | 6.59 | 5.51 | 4.28 |
| | Ensemble scheme 4 | 4.66 | 4.44 | 3.90 | 3.30 | 2.56 |
| | Ensemble scheme 5 | 5.02 | 4.37 | 3.92 | 3.30 | 2.57 |
| | Ensemble scheme 6 | 4.95 | 5.01 | 3.98 | 3.51 | 2.61 |
| Average interval score | Bayesian non-regression | 5.85 | 5.22 | 4.71 | 4.16 | 3.53 |
| | Bayesian regression | 5.91 | 5.24 | 4.73 | 4.17 | 3.54 |
| | Linear regression | 5.87 | 5.23 | 4.72 | 4.17 | 3.54 |
| | Quantile regression | 6.55 | 5.49 | 4.84 | 4.22 | 3.57 |
| | Ensemble scheme 1 | 5.96 | 5.30 | 4.77 | 4.20 | 3.56 |
| | Ensemble scheme 2 | 5.93 | 5.28 | 4.75 | 4.18 | 3.55 |
| | Ensemble scheme 3 | 9.15 | 8.06 | 7.16 | 6.19 | 5.10 |
| | Ensemble scheme 4 | 7.62 | 5.83 | 5.02 | 4.31 | 3.62 |
| | Ensemble scheme 5 | 6.43 | 5.48 | 4.84 | 4.21 | 3.56 |
| | Ensemble scheme 6 | 188.33 | 48.85 | 25.36 | 12.38 | 6.66 |

**Acknowledgements:** We sincerely thank the Editor, the Associate Editor, Dr Elena Volpi and an anonymous referee for their very detailed reviews and suggestions, which helped us to significantly improve the paper during revisions. We are also grateful to Dr Hristos Tyralis for constructive remarks and fruitful discussions.

**Declarations of interest:** The authors declare no conflict of interest.

**Funding information:** The research work of Georgia Papacharalampous was supported by the Hellenic Foundation for Research and Innovation (HFRI) under the HFRI PhD Fellowship grant (Fellowship Number: 1388).